\documentclass[twocolumn,aps,superscriptaddress,pre]{revtex4}

\usepackage{amsmath,amssymb,graphicx}
\usepackage{subfigure}
\usepackage{algorithmic,algorithm}
\usepackage{enumerate}
\usepackage{times}
\usepackage{color}
\usepackage{soul}
\allowdisplaybreaks[4]

\definecolor{yblue}{rgb}{0.06, 0.3, 0.57}
\usepackage[pdftex]{hyperref}
\hypersetup{colorlinks=true,linkcolor=blue,citecolor=blue,urlcolor=blue}

\newcommand{\state}[1]{|{#1}\rangle}
\newcommand{\statem}[1]{$|{#1}\rangle$}

\begin{document}

\title{Systematic solitary waves by linear limit continuation from two anisotropic traps in two-dimensional Bose-Einstein condensates}

\author{Wenlong Wang}
\email{wenlongcmp@scu.edu.cn}
\affiliation{College of Physics, Sichuan University, Chengdu 610065, China}

%\date{\today}

\begin{abstract}
Linear limit continuation was recently developed as a systematic and effective method for constructing numerically exact solitary waves from their respective linear limits. In this work, we apply the technique to two typical anisotropic harmonic traps in two-dimensional Bose-Einstein condensates to further establish the method and also to find more solitary waves. Many wave patterns are identified in the near-linear regime and they are subsequently continued into the Thomas-Fermi regime, and then they are further continued into the isotropic trap if possible. Finally, the parametric connectivity of the pertinent solitary waves is also discussed.
\end{abstract}

\maketitle

\section{Introduction}
Bose-Einstein condensates (BECs) have provided an excellent platform for investigating (vector) solitary waves \cite{becbook1,becbook2,Panos:book}. The interatomic interactions, the trap potential, and the spin state can be flexibly tuned, and the mean-field theory is frequently relevant. In the most common repulsive condensates, a diverse set of solitary waves have been extensively studied, e.g., the dark solitons \cite{Dimitri:DS}, vortices, vortex rings \cite{Ionut:VR,Wang:VR,Wang:VRB}, and even knots \cite{Ruban:Knots}. Vector solitary waves can exhibit intriguing properties, e.g., a massive vortex may feature unusual flowering trajectories and instabilities compared with a regular vortex \cite{Andrea:VB,Jennie:VB}. Interestingly, many of these structures are also relevant in a much broader setting, e.g., in normal fluids, superfluids and superconductors, and nonlinear optics \cite{Nicolay:SF,DSoptics}.

Solitary wave solutions are of paramount importance for exploring the existence, stability, dynamics, and interactions of various solitary waves. While analytic methods are highly interesting, they are essentially limited to the 1D homogeneous integrable setting \cite{Lichen:DT}, and some particular solutions in special settings. As such, it is an outstanding research track to find numerically exact solitary wave solutions, particularly in nonintegrable systems. %They are important for investigating the pattern formation of the solitary waves.
They offer insights into both static and dynamic properties, and the latter can be investigated via either the Bogoliubov-de Gennes linear stability analysis or direct time evolution.
%the Bogoliubov-de Gennes (BdG) linear stability analysis
%As such, both robust oscillations and symmetry-breaking instabilities of a solitary wave can be studied in detail. 
Furthermore, this field is computationally interesting in its own right, and interesting algorithms and their properties are extensively explored.
%In a generic nonintegrable setting, finding numerically exact solutions is highly interesting.
%To study solitary wave properties, it is of paramount importance to find (numerically) exact solutions. 
%One major theme of research is finding... to study their properties.
A pioneering technique to find numerically exact states at the systematic level is the deflation method \cite{Panos:DC1,Panos:DC2,Panos:DC3}. This method keeps track of a list of found solutions, and then avoids them in further search by suitably modifying the pertinent equations. 

The linear limit continuation (LLC) was recently developed as an alternative method for systematically constructing numerically exact solitary waves from their respective linear limits \cite{Wang:DD,Wang:MDDD,Wang:DAD,Wang:LLC}. This method appears to be very effective, yet it is technically much simpler \cite{Wang:LLC}. It also offers a theoretical framework to organize the obtained waves, and provides insights into their existence by tracing back to their linear counterparts. It should be noted that the idea of constructing solitary waves from their linear limits itself has been intensively employed in earlier works to study particular waveforms by physical insights \cite{Carr:VX,Wang:DSS,Wang:VR}. 
Here, we systematically find solitary waves, focusing on the 2D scaler field with repulsive interactions \cite{Wang:LLC}. In this setting, the fundamental solitary wave excitations are dark soliton filaments and point vortices. The LLC method first classifies linear degenerate sets, which can be visualized with ``lattice planes'' \cite{Wang:LLC}. Then, we work with the linear degenerate sets in turn. The central idea of the LLC method is very simple, it finds distinct wave patterns bifurcating from a linear degenerate set by making a suitable linear combination of the pertinent linear degenerate states in the near-linear regime. This is carried out by a random solver, following the framework of the (degenerate) perturbation theory. Next, each identified wave pattern is continued into the nonlinear Thomas-Fermi (TF) regime by a numerical continuation in the chemical potential. Similarly, a continuation in the trap is also frequently relevant. Naturally, the classification of the linear degenerate sets depends on the trap geometry. The method was applied to the isotropic trap and a prototypical anisotropic trap of aspect ratio $1/2$, finding a spectacular set of low-lying wave patterns \cite{Wang:LLC}. However, the method has not been applied to other anisotropic traps.

The main purpose of this work is to apply the LLC method to two additional anisotropic traps of prototypical aspect ratios $1/3$ and $2/3$, respectively. Here, we have two main motivations. First, we aim to further establish the LLC method. This is especially important as the method was recently developed, and it is sensible to examine its effectiveness and robustness before extending it further to the three-dimensional setting and multi-component systems. Second, it is also interesting to identify more solitary waves in this exploration. %This is helpful for investigating their bifurcation relations \cite{Panos:DC1,Panos:DC2,Panos:DC3,Brand_SV}, and for studying the pattern formation. 
As more wave patterns are found, we aim to examine the parametric connectivity of the pertinent solitary waves. Indeed, a solitary wave may bifurcate from different traps, then it is important to confirm whether or not the relevant similar-looking states continued from different paths are the same state, to our knowledge. This is a nontrivial task, as two similar states are not necessarily identical, and conversely two apparently different states may be parametrically connected, showing the rather intricate nature of the bifurcation processes of solitary waves. In addition, the understanding of 2D waveforms is helpful for the exploration of 3D wave patterns, because some 3D states are essentially extensions of the 2D counterparts along the $z$ axis. Our investigation is largely successful. While some repetitive states are obtained, many novel wave patterns are found as expected, showing that LLC is an effective method for systematically constructing solitary waves.

This work is organized as follows. We present the theoretical and numerical setup in Sec.~\ref{setup}. The method is applied to two anisotropic traps, and the continuation of the identified low-lying solitary waves is illustrated and discussed in Sec.~\ref{results}. Finally, a summary and an outlook of future directions are given in Sec.~\ref{co}.

\section{Theoretical and numerical setup}
\label{setup}

We study the following dimensionless Gross-Pitaevskii equation in two dimensions:
\begin{align}
    -\frac{1}{2} \nabla^2 \psi+V \psi +| \psi |^2 \psi = i \frac{\partial \psi}{\partial t},
    \label{GPE}
\end{align}
where $\psi(x,y,t)$ is the macroscopic wavefunction, $V=(\omega_x^2 x^2 + \omega_y^2y^2)/2$ is the harmonic potential. We set $\omega_y=1$ for convenience by scaling without loss of generality, and $\kappa=\omega_y/\omega_x$ is the trap aspect ratio. Stationary state of the form $\psi(x,y,t)=\psi^0(x,y) e^{-i\mu t}$ leads to:
\begin{align}
    -\frac{1}{2} \nabla^2 \psi^0 +V \psi^0 +| \psi^0 |^2 \psi^0 = \mu \psi^0,
    \label{GPEstationary}
\end{align}
where $\mu$ is the chemical potential. This is the central equation we aim to solve in this work. This equation has a linear limit, the two-dimensional quantum harmonic oscillator, when the field has a vanishing amplitude. In this setting, the normalized linear state in Cartesian coordinates reads:
\begin{align}
\varphi_{n_x,n_y}^0(x,y) = \frac{H_{n_x}(\sqrt{\omega_x}x)H_{n_y}(\sqrt{\omega_y}y)}{\sqrt{\pi2^{n_x}n_x!2^{n_y}n_y!}\kappa^{1/4}} e^{-(\omega_x x^2 + \omega_y y^2)/2}. \label{linearsc}
\end{align}
Here, $H$ is the Hermite polynomial, and the corresponding eigenenergy is $\mu_0=(n_x+1/2)\omega_x+(n_y+1/2)\omega_y$, where $n_x, n_y = 0, 1, 2, ...$ The two quantum numbers represent the number of linear dark nodes along the $x$ and $y$ directions, respectively.

\begin{figure}%{r}{0.5\textwidth}
\includegraphics[width=0.495\columnwidth]{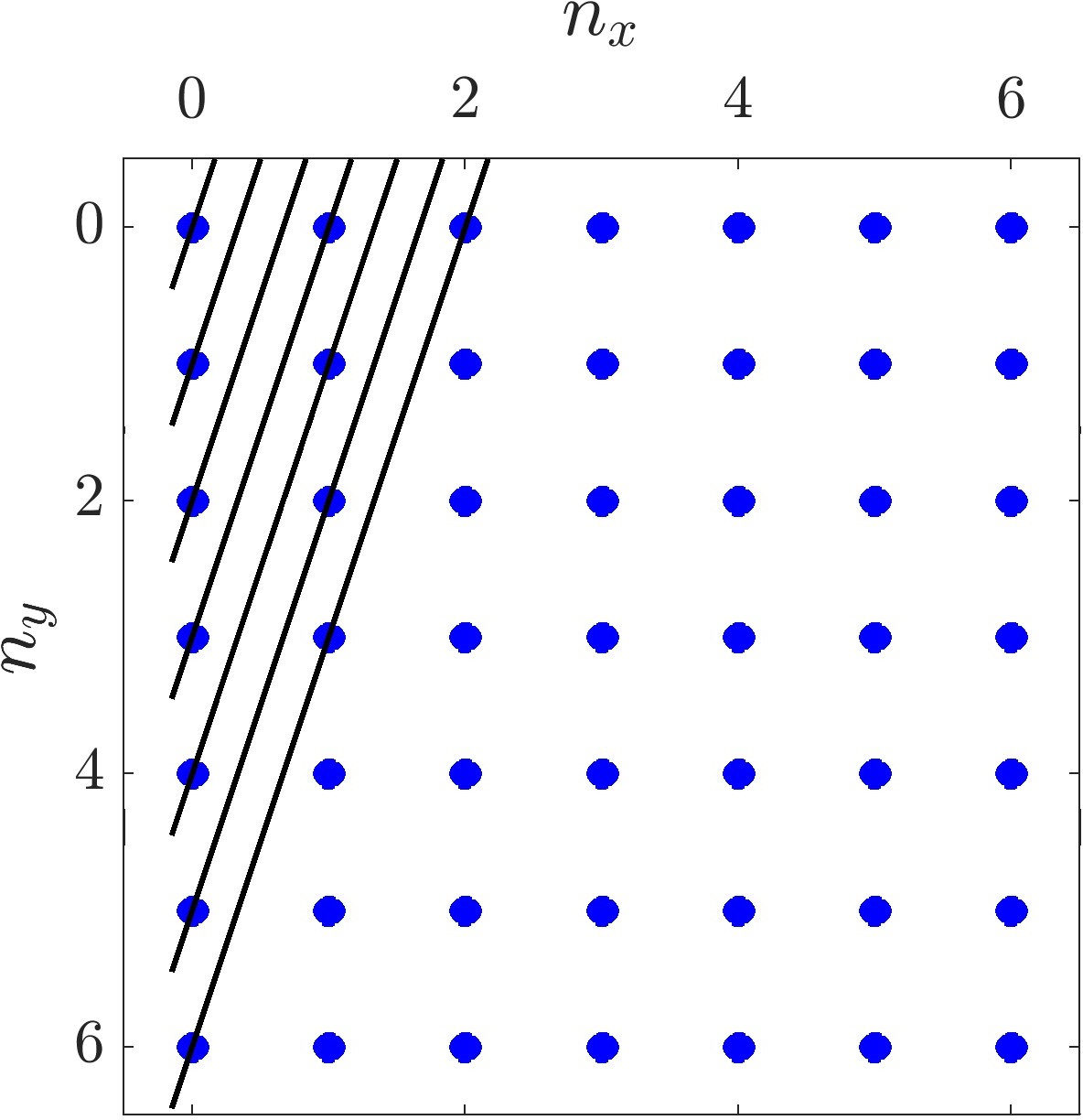}
\includegraphics[width=0.495\columnwidth]{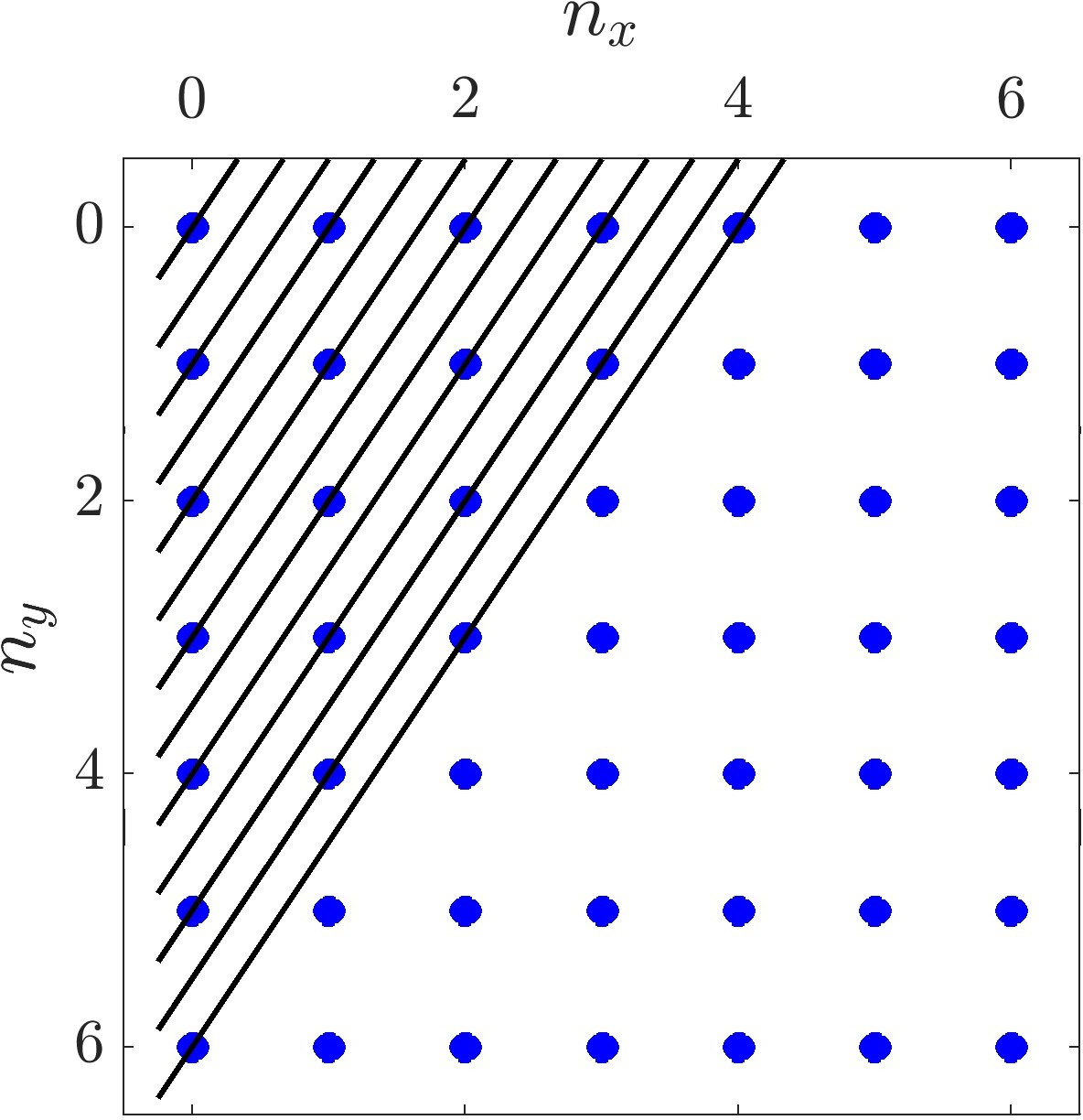}
%\put (-225,183) {(a) $\lambda=0.5$}
%\put (-145,183) {(b) $\lambda=1$}
%\put (-70,183) {(c) $\lambda=3$}
\caption{Linear degenerate sets can be geometrically represented by lattice planes, focusing only on the $n_x, n_y \geq 0$ region. Here, the low-lying degenerate sets of two prototypical sets of lattice planes $[3, 1]$ (left panel) and $[3, 2]$ (right panel) are illustrated, and the two scenarios correspond to trap aspect ratios $\kappa=1/3$ and $2/3$, respectively.
In general, the lattice plane $[p, q]$ corresponds to a trap aspect ratio $\kappa=q/p \leq 1$.
}
\label{XY}
\end{figure}

The LLC starts by classifying the different linear degenerate sets, which depend on the trap aspect ratio. Interestingly, they can be geometrically represented by the ``lattice planes'' of the $(n_x, n_y)$ quantum numbers. The different sets of lattice planes correspond to different trap aspect ratios. We systematically examined the low-lying linear degenerate sets of $\kappa=1$ and $\kappa=1/2$ in a previous work \cite{Wang:LLC}. Here, we study the low-lying linear degenerate sets of $\kappa=1/3$ and $\kappa=2/3$, as illustrated in Fig.~\ref{XY}. The second lattice plane of $\kappa=2/3$ strikes through no lattice point in our restricted $n_x, n_y \geq 0$ region. Naturally, this lattice plane is skipped in our setup, and the other lattice planes yield physical linear degenerate sets for investigation.

The LLC method treats the linear degenerate sets in turn. For a generic and finite linear degenerate set, only a finite number of wave patterns bifurcate from the set, in line with the degenerate perturbation theory. Consider a set of $K$ orthonormal linear degenerate states $\varphi_i^0, \ i=1, 2, ..., K$ with eigenenergy $\mu_0$. The degenerate perturbation analysis works as:
\begin{align}
    \varphi^0 &=\sum_i c_i \varphi_i^0, \quad \sum_i|c_i|^2=1, \label{norm} \\
    \psi^0 &=\sqrt{\epsilon}\varphi^0, \label{scaling} \\
    \mu &= \mu_0 + \epsilon \mu_r, \label{eps} \\
    \mu_r &= \iint |\varphi^0|^4 dx dy, \label{mur} \\
    \mu_rc_i &=\iint \varphi^2\varphi^*\varphi_i^*dxdy. \label{murci}
\end{align}
Here, $\epsilon$ is an arbitrary small perturbation parameter and the stationary superscript for the fields in the last equation is omitted for clarity. If we multiply the final equation by $c_i^*$ and then sum over $i$, we obtain the formula of $\mu_r$, therefore, Eq.~\eqref{mur} is not an independent constraint. Nevertheless, it is very important for our numerical setup. It is possible to calculate these coefficients theoretically, see, e.g., the Lyapunov–Schmidt reduction technique, however, this is frequently rather tedious to apply even to low-lying states \cite{RCG:rotating}. 
In our setup, we do not directly solve for the unknowns, they are instead calculated by a random solver, as we should discuss below. It is evident that only certain linear combinations correspond to nonlinear stationary states because of the constraints, the nonlinear locking phenomenon. The linear state $\varphi^0$ is defined as the underlying linear state (ULS) of the nonlinear wave $\psi^0$ in the near-linear regime. The significance of the ULS is that $\sqrt{\epsilon}\varphi^0$ is a good representation of the nonlinear wave $\psi^0$ at $\mu=\mu_0+\epsilon\mu_r$ to leading order in a reasonable interval in the near-linear regime.

The LLC method finds distinct wave patterns by a random solver. Here, we briefly describe the method, and we refer the interested readers to \cite{Wang:LLC} for details. For a generic linear degenerate set of eigenenergy $\mu_0$, we work at a chemical potential $\mu \gtrsim \mu_0$ in the near-linear regime. First, we prepare a random initial guess by a random linear combination of the linear degenerate states. This linear state is then suitably normalized as $\varphi$ as in Eq.~\eqref{norm}, this is not $\varphi^0$ yet as it does not necessarily correspond to a stationary nonlinear wave. Next, the linear state $\varphi$ is properly scaled to obtain a nonlinear state $\psi$ following the degenerate perturbation theory. We calculate the $\mu_r$ for the $\varphi$ field as in Eq.~\eqref{mur}, and then find $\epsilon$ for our working chemical potential by Eq.~\eqref{eps}, and subsequently $\varphi$ is scaled to a nonlinear field following Eq.~\eqref{scaling}.
%the scaling factor depends on the waveform and the working chemical potential. 
Then, this nonlinear state $\psi$ is applied to the Newton's solver for convergence to find the numerically exact stationary state $\psi^0$. If converges, we can normalize the state to find $\varphi^0$, which is an ULS. Finally, this ULS is projected to the basis states to calculate the
%An expansion of the ULS on the linear degenerate set yields the 
numerically exact linear combination coefficients of Eq.~\eqref{norm}. The art of the random solver is that we can obtain numerically exact linear combination coefficients, despite that the initial linear combination coefficients are not exact. 
This process is repeated many times to identify all the distinct wave patterns and their ULSs in the low-density regime. To identify distinct solitary waves, we examine a few statistics of the nonlinear waves, e.g., the norm $\iint |\psi^0|^2dxdy$, the maximum magnitude $|\psi^0|_{\rm{max}}$, and the central magnitude $|\psi(0)|$. Note that they are invariant upto symmetry transformations, and the statistics are also relevant for comparing wave patterns beyond the near-linear regime. 

After identifying the distinct wave patterns, we continue them into the TF regime by a numerical continuation. In the near-linear regime, we prepare an initial guess following the perturbation theory %for a nonlinear wave by a suitable scaling of its ULS 
for a few steps, because the field magnitude typically grows pretty rapidly in the vicinity of the linear limit. 
When the field reaches a moderate density, the continuation can run on its own, i.e., a converged solution can be used as the initial guess of the next chemical potential, and so on. %In this way, the distinct solitary wave patterns are continued from the near-linear regime into the TF regime. 
For a solitary wave in the TF regime, one can frequently continue it further in the trap frequency $\omega_x$. Indeed, this $\omega_x$ is also a perfectly valid parameter from the perspective of Eq.~\eqref{GPEstationary}. In this work, we also try to continue the solitary waves into the isotropic trap of $\omega_x=1$ if possible.

%In practice, the ULS are found by a random solver. We refer the interested readers to \cite{Wang:LLC} for details and here we briefly describe the idea. First, a normalized linear state is prepared by an appropriate random linear combination of the linear degenerate states. Next, it is suitably scaled following the perturbation theory at a working $\mu \gtrsim \mu_0$, and it is applied as the initial guess for the Newton's solver for convergence. If the search converges, the numerically exact solution and its ULS are found. The ULS is then expanded on the basis states to calculate the numerically exact linear combination coefficients. The process is repeated for $M \sim O(1000)$ times to find all the distinct wave patterns. These wave patterns are then continued in the chemical potential into the TF regime.

We also investigate the parametric connectivity of the solitary waves found when relevant in this work. Here, we illustrate the idea with an example. We obtained the DS01 state (a dark soliton stripe from the \statem{01} linear state) at $\mu=16$ and $\kappa=1/2$ in \cite{Wang:LLC}, which is also continued into the isotropic trap of $\omega_x=1$. This dark soliton stripe is also found here at $\mu=16$ and $\kappa=1/3$, which is similarly continued into the isotropic trap. It is a natural question whether the final states are identical. In fact, the two continuation paths cross and then overlap in the parameter space, i.e., when $\omega_x$ is decreased from $3$ to $1$, it passes the intermediate value $2$ in the continuation. As such, it is even more interesting to check whether the states are already identical when their continuation paths first cross. To this end, we compare their statistics. If all of their statistics agree to high accuracy, then the states are likely identical. Indeed, we have checked that the dark soliton stripe states above are identical as expected, and its continuation appears to be path independent, to our knowledge. We frequently do such crosschecks in our work when relevant, such that we gain a better understanding of their continuation in the parameter space. It should be emphasized that this is far from a trivial task, e.g., the DS02 states (two dark soliton stripes from the \statem{02} linear state) of $\kappa=1/2$ and $\kappa=1/3$ are parametrically connected but they are then parametrically connected with the ring dark soliton (RDS) of the isotropic trap. This shows the intricate nature of the continuation and bifurcation of solitary waves.

Finally, we present the numerical methods and typical simulation parameters. Our field is discretized with a square lattice utilizing the finite element method. The computational domain is $[-8, 8]$ with a step size of $h=0.04$ in each spatial direction, and zero boundary condition is applied as the field is bounded. The pertinent nonlinear equations are subsequently solved with the Newton's method, converting a nonlinear problem to a series of linear ones. 
We adopt a piecewise constant continuation schedule for both $\delta \mu$ and $\delta \omega_x$ for simplicity, and typical values are $0.005$ or smaller for both parameters. 
If a continuation step fails and requires a finer step, we restart the continuation from a successfully converged state with a smaller step size. 
%To benchmark the simulation parameters, we have carefully continued the pertinent polar states in both the $1d$ and the $2d$ frameworks, requiring that the results are consistent within the numerical accuracy.
Next, we discuss the numerical differentiation schemes in detail because we quickly encounter linear states of high quantum numbers, e.g., the linear state \statem{06} for both traps. The computation of such excited states requires some attention. %as a numerical setup that works well for relatively low-lying states may no longer be fully reliable for such highly excited states. 
%Numerical simulation of solitary waves of highly excited states in the TF regime has not been studied very much, to our knowledge. Here, we examine a few estimators and propose a sixth-order estimator of the Laplacian operator for benchmarking the finite element method. 
In an earlier work, we found that the regular $5$-point Laplacian is not sufficiently accurate for a typical spacing $h$ when the waves are moderately excited \cite{Wang:LLC}. As such, we implement here the square-shaped $9$-point Laplacian, which works quite well at least upto the quantum number $4$ \cite{Wang:LLC}.
%However, we encounter states from quite highly excited states with quantum number $6$ in this work, and it is important to examine the accuracy of the present scheme. 
To examine its accuracy for more excited states, we studied various polar states \cite{Wang:LLC}, and also some other waveforms. 
%as the ring dark solitons deform when the simulation is not sufficient accuracy
We find that this Laplacian appears to work well upto the quantum number $5$ but may not be fully reliable for the quantum number $6$ with the present parameters. Particularly, the RDS3 suffers a deformation, but $h=0.03$ can fix the problem. 
To make sure the results are reliable, we also implemented a genuinely fourth-order estimator. It is a plus-shaped $9$-point Laplacian. The one-dimensional counterpart is:
\begin{align}
f''(x) = \frac{-30f(x)+16f(x\pm h)-f(x\pm 2h)}{12h^2} + O(h^4).
\end{align}
Here, we assume that we sum over all the separate terms.
The 2D estimator is obtained by a summation of the 1D estimators applied along the $x$ and $y$ directions, respectively. This estimator is clearly fourth-order because no mixed derivatives arise in the Taylor expansion. By contrast, the square-shaped $9$-point Laplacian is not a genuinely fourth-order estimator, it is designed to be fourth-order for the Laplacian equation. It seems that this genuinely fourth-order estimator does work better than the square-shaped quasi-fourth-order estimator for the present problem, and both are $9$-point estimators. In our various experiments, they yield essentially the same results with $h=0.04$ and $h=0.03$, respectively. Therefore, we implement the genuinely fourth-order estimator for the linear degenerate sets with a quantum number $6$ in this work. A similar extension to a plus-shaped six-order estimator is also possible, it seems highly accurate but considerably slower, at least for a direct solution of the linear equations. It appears to work well even for the rather excited RDS4 and RDS5 states at the level of $h=0.04$. Since we do not study such excited states in this work, we should not discuss this further. 
To understand the nature of solitary waves and their evolution, we analyze their density and phase profiles, field contours, and their statistics. In contour analysis, we examine the contours of Re$(\psi^0)=0$ and Im$(\psi^0)=0$. This plays a crucial role in understanding the nature of certain complicated wave patterns, e.g., with strong density depletion regions or overlapping vortices.

\section{Results}
\label{results}

Here, we study the low-lying states bifurcated from the linear degenerate sets of two anisotropic traps of $\kappa=1/3$ and $2/3$, respectively. We discuss the wave patterns, their chemical and trap continuation, and the parametric connectivity of the pertinent states. Note that the low-lying solitary waves of the isotropic trap $\kappa=1$ and the anisotropic trap $\kappa=1/2$ are systematically investigated in \cite{Wang:LLC}.

\subsection{Wave patterns from trap aspect ratio 1/3}

\begin{figure*}[t]%{r}{0.5\textwidth}
\includegraphics[width=\textwidth]{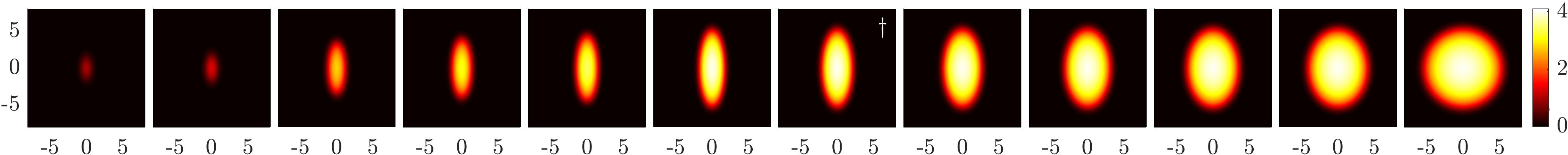}
\includegraphics[width=\textwidth]{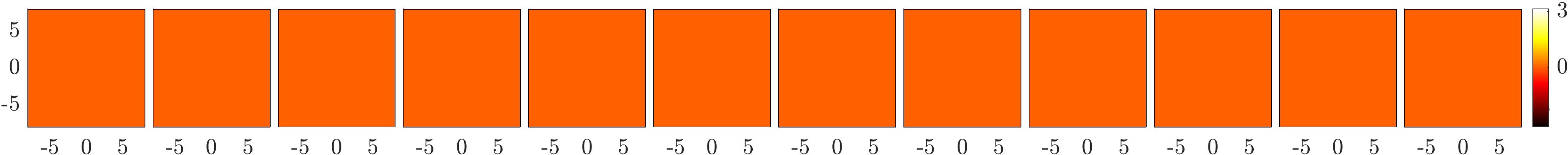}
\includegraphics[width=\textwidth]{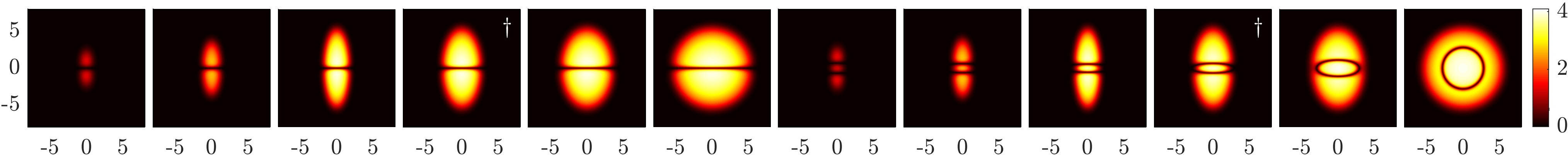}
\includegraphics[width=\textwidth]{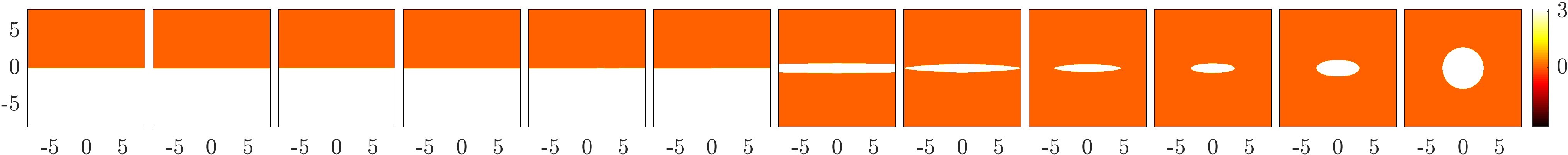}
\includegraphics[width=\textwidth]{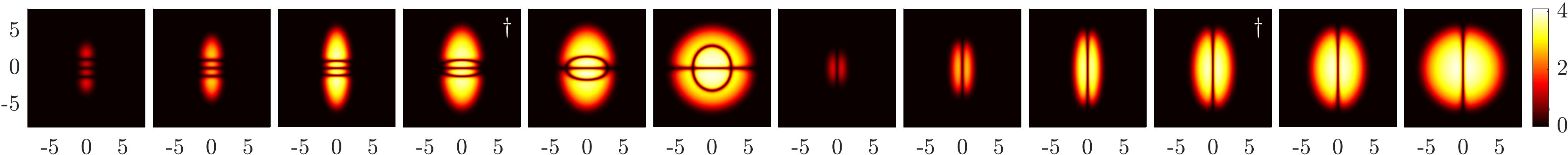}
\includegraphics[width=\textwidth]{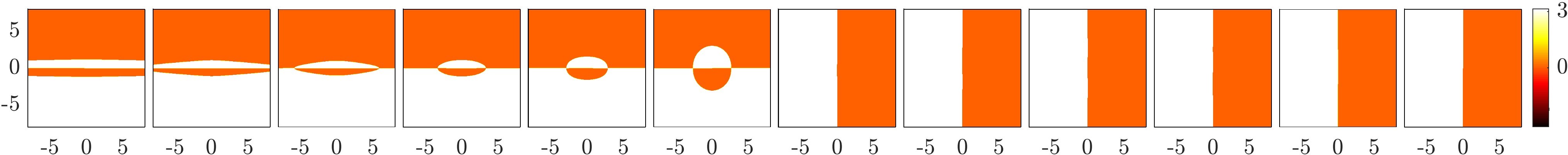}
\includegraphics[width=\textwidth]{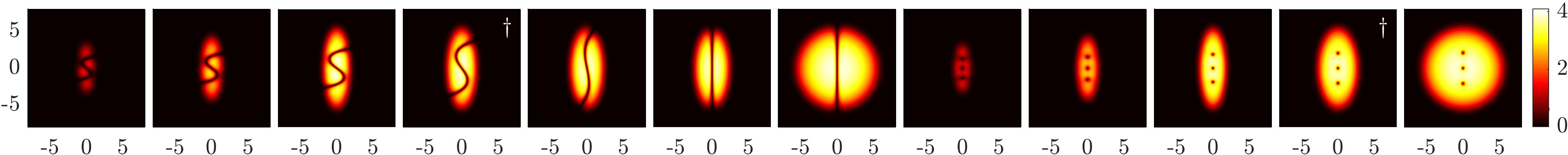}
\includegraphics[width=\textwidth]{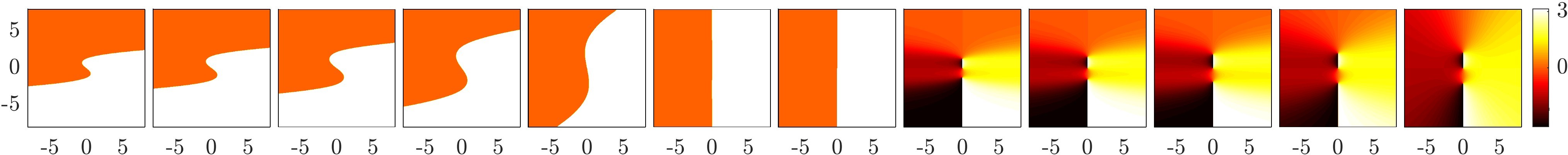}
%\put (-225,183) {(a) $\lambda=0.5$}
%\put (-145,183) {(b) $\lambda=1$}
%\put (-70,183) {(c) $\lambda=3$}
\caption{
Solitary waves continued from the lowest four $(\mu_0=2, 3, 4, 5)$ linear degenerate sets of $\kappa=1/3$. The first set shows the ground state at $\mu=2.5, 3, 8, 10, 12, 16$ of $\omega_x=3$, followed by a further continuation in $\kappa$ at fixed $\mu=16$. Here, the states at $\omega_x=2.5, 2, 1.7, 1.5, 1.3, 1$ are depicted, and the dagger symbol highlights the onset of the trap continuation. The second set depicts the DS01 state at $\mu=4, 8, 16$ and $\omega_x=2, 1.5, 1$, and the DS02 state at $\mu=5, 9, 16$ and $\omega_x=2, 1.5, 1$. The third set illustrates the DS03 and DS10 states at $\mu=6, 10, 16$ and $\omega_x=2, 1.5, 1$. The fourth set sketches the S soliton and the VX3 state at $\mu=6, 10, 16$ and $\omega_x=2.7, 2.4, 2.3, 1$, and $\omega_x=2, 1$, respectively.
}
\label{GSa}
\end{figure*}

The linear excitation spectrum is $\epsilon_{mn}=(m+1/2)/\kappa+(n+1/2)=3m+n+2$, where $\kappa=1/3$ and $m, n=0, 1, 2, ...$ The ground state \statem{00}, the first excited state \statem{01}, and the second excited state \statem{02} have $\mu_0=2, 3, 4$, respectively, and they are nondegenerate in this trap. Naturally, their respective ULSs are:
\begin{align}
    \varphi_{\mathrm{GS}}^0 &= \state{00}, \quad \mu_0=2, \\
    \varphi_{\mathrm{DS01}}^0 &= \state{01}, \quad \mu_0=3, \\
    \varphi_{\mathrm{DS02}}^0 &= \state{02}, \quad \mu_0=4.
\end{align}
The chemical potential and trap continuations of these states are quite robust, as illustrated in Fig.~\ref{GSa}. They are also similar to their corresponding $\kappa=1/2$ counterparts, including their evolutions. As expected, the \statem{00} state is continued into the TF ground state. It has a uniform phase, providing the background for hosting solitary waves. It approximately takes a profile of $\psi^0_{\mathrm{TF}}=\sqrt{\max(\mu-V, 0)}$. The \statem{01} state becomes the DS01 state with a horizontal dark soliton stripe in the center. It features a density node and a phase jump of $\pi$ crossing the node. We mention in passing that the trap continuation is also relevant dynamically, e.g., a recent work shows that a dark soliton stripe can generate a single vortex (rather than a vortex dipole) in an adiabatic continuation of the trap frequency \cite{Lichen:DSVX}.

We have checked carefully that the ground states of $\kappa=1, 1/2, 1/3$ are all parametrically connected, to our knowledge. This means that when the continuations meet in the parameter space, the same state is obtained, i.e., the continuation is path independent.
For example, we can directly continue the ground state to $\mu=16$ at $\kappa=1/2$ or alternatively we can first continue it to $\mu=16$ at $\kappa=1/3$ and then continue it further to $\kappa=1/2$, the obtained states are identical. The same is true for the DS01 states. It seems likely that these two states are ULSs for all the harmonic traps and the corresponding states are parametrically connected. 

%The continuation of the DS02 state is as expected, but remains somewhat counterintuitive. As the chemical potential increases, the two dark soliton filaments connect into a closed loop in the TF regime. As it is continued into the isotropic trap, it gradually morphs into the polar RDS. 

The DS02 state has two horizontal dark soliton stripes in the near-linear regime, however, they tend to curve towards each other rather than towards the edge of the condensate as the chemical potential increases. As such, they quickly connect into a single closed loop in the chemical potential evolution, see, e.g., the phase profile of $\mu=9$. It is remarkable that when it is continued into the isotropic trap, it gradually morphs into the polar RDS state rather than the regular DS02 state of the isotropic trap \cite{Wang:LLC}. In this process, a series of elliptical RDSs is obtained. This transformation appears to be highly robust, e.g., it consistently happens at very low chemical potentials of $\mu=4.5, 4.3, 4.1$. To our knowledge, the DS02 states of $\kappa=1/2, 1/3$ are parametrically connected, and they are parametrically connected to the RDS of the isotropic trap \cite{Wang:LLC}.

The fourth linear degenerate set has two states \statem{03} and \statem{10} with eigenenergy $\mu_0=5$. We have identified four wave patterns, and the ULSs are summarized here:
\begin{align}
    \varphi_{\mathrm{DS03}}^0 &= \state{03}, \\
    \varphi_{\mathrm{DS10}}^0 &= \state{10}, \\
    \varphi_{\mathrm{S}}^0 &= 0.8492\state{03}-0.5280\state{10}, \\
    \varphi_{\mathrm{VX3}}^0 &= 0.7642\state{03}+0.6449i\state{10}.
\end{align}
These coefficients are estimated using the numerically exact states at $\mu=5.04$, but they depend only very weakly on $\mu$ in the near-linear regime. The continuation of these states is illustrated in the third and fourth sets of Fig.~\ref{GSa}. The DS03 state has three dark soliton stripes in the near-linear regime, and the two outer ones bend towards each other as the chemical potential increases, therefore, the state evolves into the $\phi$ soliton in the TF regime. This is more evident when it is continued into the isotropic trap. This process is very similar to that of the above DS02 state, except that there is an additional horizontal dark soliton stripe. Similarly, the transformation is quite robust, and it persists down to the very low chemical potentials of $\mu=5.5, 5.3, 5.1$. The DS03 states of $\kappa=1/3, 1/2$ are parametrically connected, to our knowledge, but both states morph into the $\phi$ soliton in the isotropic trap. The $\phi$ soliton itself has a linear limit in the isotropic trap \cite{Wang:LLC}.

The DS10 state differs from the DS01 state here as the former dark soliton is excited along the hard direction and the latter dark soliton is excited along the soft direction of the condensate. The continuation of the DS10 state in both the chemcial potential and the trap is very robust, and it becomes the regular DS10 state in the isotropic trap in the TF regime. The DS10 state and the DS01 state above converge to the same state in the isotropic trap, to our knowledge, they become a vertical and a horizontal dark soliton stripe, respectively.

The S soliton has a single dark soliton filament with an interesting curving structure, and it has a well-defined odd parity inherited from the linear states. It is like two U solitons linked together \cite{Wang:LLC}. Similarly, the U soliton is from a linear combination of the $\state{02}, \state{10}$ states in the $\kappa=1/2$ trap. The ULSs enable us to extract the positions of the dark soliton filaments by setting the ULSs to $0$. It is evident that $x \propto H_3(y)$ and $x \propto H_2(y)$ for the two filaments. Precisely, we calculate that the curves are approximately described by $x=0.3791(2y^3-3y)$ and $x=0.5419(2y^2-1)$ for the S and U solitons, respectively. 
%The ULSs offer insights into the emergence of dark soliton filaments of such shapes. 
Furthermore, we can appreciate the existence of U-shaped and S-shaped vortical filaments in 3D \cite{Ionut:VR03}, they can arise from a complex mixing of such extended U-shaped or S-shaped surfaces and other intersecting dark soliton surfaces. As the relevant linear states get more excited, or higher order polynomials are involved, more complicated dark soliton filaments in 2D and vortical filaments in 3D can emerge. 
The continuation in the chemical potential is pretty robust. As $\omega_x$ decreases, the S soliton gradually opens a larger angle like the U and $\Psi$ solitons \cite{Wang:LLC}. Then, it morphs into the dark soliton stripe DS10 state at $\omega_x \lesssim 2.4$. 

The aligned vortex triple VX3 state has three vortices of alternating charge, as it is from a complex mixing of the \statem{03} and \statem{10} states. The continuation in both the chemical potential and the trap is highly robust. The VX3 has no linear limit in the isotropic trap as the two basis states \statem{03} and \statem{10} are not degenerate therein. Numerical continuation confirms this, if we gradually decrease the chemical potential in the isotropic trap, we find a lower critical chemical potential $\mu_c \approx 4.91$, in line with \cite{PK:DSVX}. The aligned vortex triple is dynamically unstable in the isotropic trap \cite{PK:DSVX}. 

It is interesting to briefly discuss the series of $n$ aligned vortices of alternating charge. In the isotropic trap, we find the single vortex state from a complex mixing of the \statem{10} and \statem{01} states. Then, the vortex dipole state emerges from a complex mixing of the \statem{10} and \statem{02} states in the $\kappa=1/2$ trap \cite{Wang:LLC}. Here, the VX3 state is obtained from a complex mixing of the \statem{10} and \statem{03} states. We have checked that the pattern continues, when \statem{10} is in turn complex mixed with \statem{04}, \statem{05}, \statem{06} in the traps of $\kappa=1/4, 1/5, 1/6$, respectively. Precisely, their ULSs are:
\begin{align}
    \varphi_{\mathrm{VX4}}^0 &= 0.7795\state{04}+0.6264i\state{10}, \quad \kappa=1/4, \label{AVQ} \\
    \varphi_{\mathrm{VX5}}^0 &= 0.7906\state{05}+0.6123i\state{10}, \quad \kappa=1/5, \\
    \varphi_{\mathrm{VX6}}^0 &= 0.7994\state{06}+0.6007i\state{10}, \quad \kappa=1/6.
\end{align}
These coefficients are computed at $\mu=1.008\mu_0$, where $\mu_0=6.5, 8, 9.5$, respectively. Their chemical potential evolution and the subsequent continuation into the isotropic trap is robust. Some of these vortical states are systematically studied in \cite{PK:DSVX}.

\begin{figure*}%{r}{0.5\textwidth}
\includegraphics[width=\textwidth]{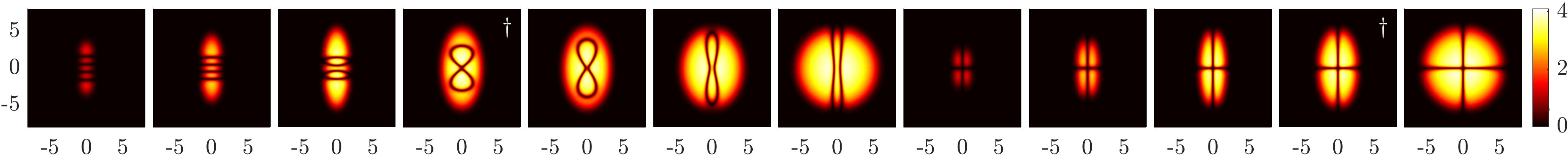}
\includegraphics[width=\textwidth]{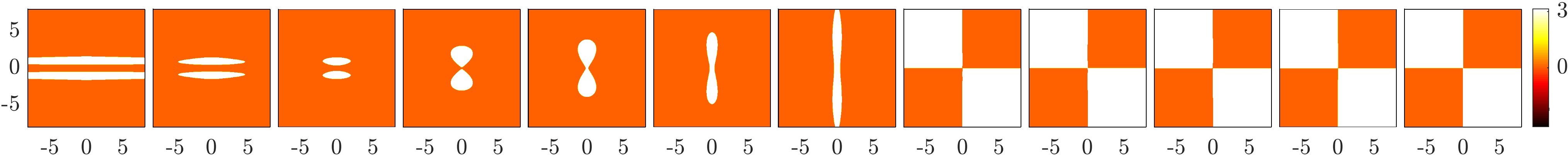}
\includegraphics[width=\textwidth]{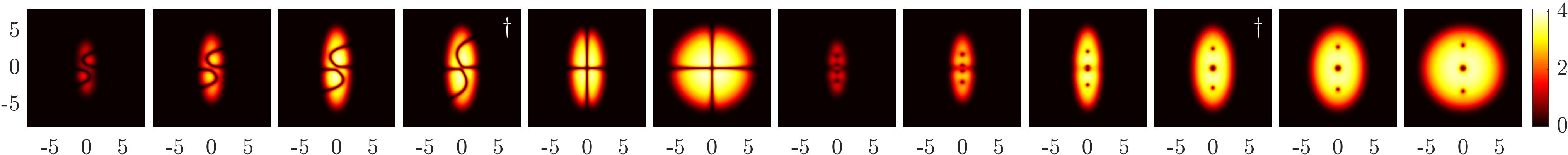}
\includegraphics[width=\textwidth]{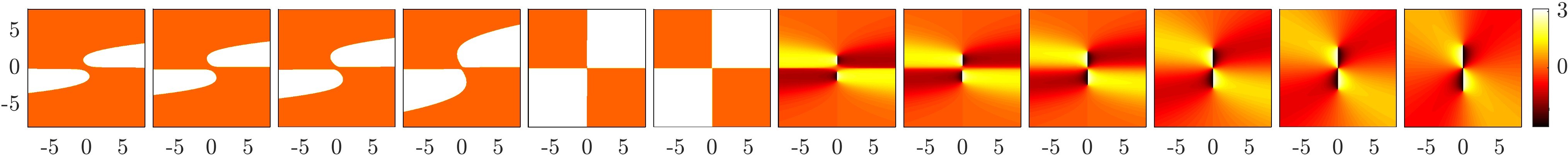}
\includegraphics[width=\textwidth]{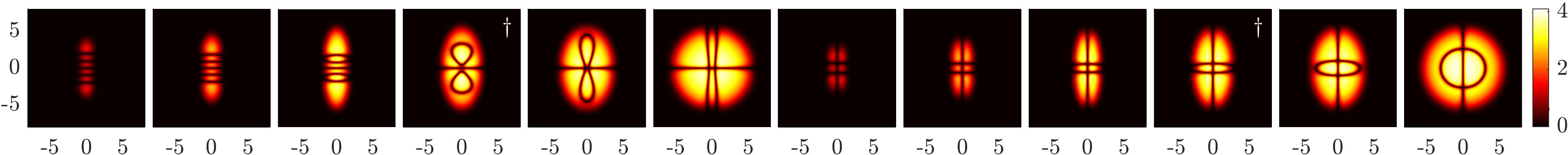}
\includegraphics[width=\textwidth]{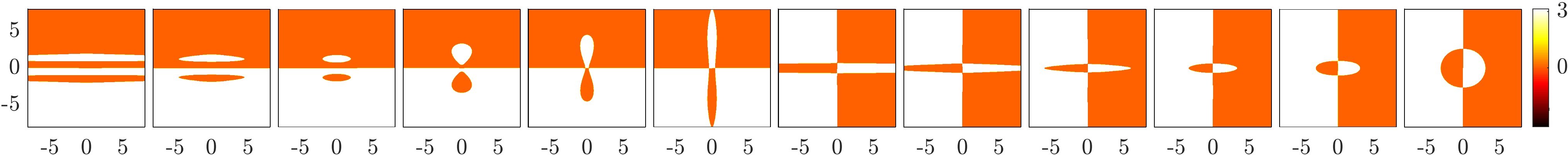}
\includegraphics[width=\textwidth]{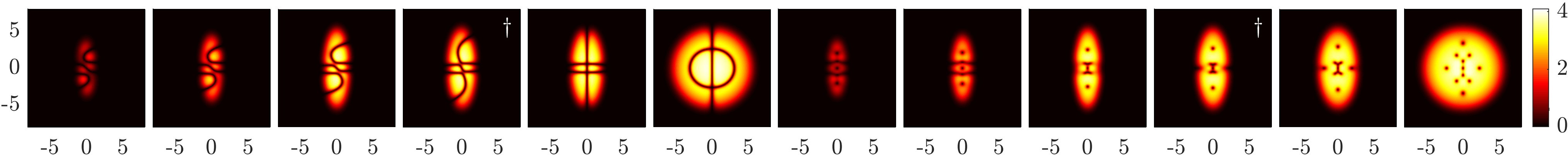}
\includegraphics[width=\textwidth]{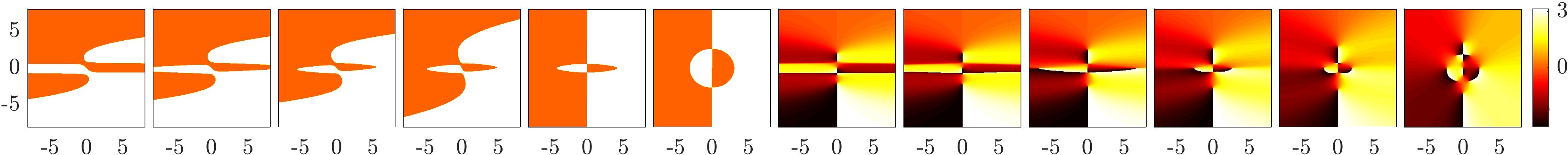}
%\put (-225,183) {(a) $\lambda=0.5$}
%\put (-145,183) {(b) $\lambda=1$}
%\put (-70,183) {(c) $\lambda=3$}
\caption{
Solitary waves continued from the $\mu_0=6, 7$ linear degenerate sets of the $\kappa=1/3$ trap. The first set shows the DS04 state at $\mu=7, 11, 16$ and $\omega_x=2, 1.7, 1.3, 1$, and the DS11 state at $\mu=7, 11, 16$ and $\omega_x=2, 1$. The second set depicts the U2 soliton and the VX4 state at $\mu=7, 11, 16$ and $\omega_x=2.7, 2.4, 1$, and $\omega_x=2, 1.5, 1$, respectively. The third set illustrates the DS05 and the DS12 states at $\mu=8, 11, 16$ and $\omega_x=2, 1.5, 1$. The fourth set sketches the U2I soliton and the VX5 state at $\mu=8, 11, 16$ and $\omega_x=2.7, 2.4, 1$, and $\omega_x=2.5, 2, 1$, respectively.
}
\label{DS04a}
\end{figure*}

The fifth linear degenerate set has two states \statem{04} and \statem{11} with $\mu_0=6$. The ULSs of the identified solitary waves are:
\begin{align}
    \varphi_{\mathrm{DS04}}^0 &= \state{04}, \\
    \varphi_{\mathrm{DS11}}^0 &= \state{11}, \\
    \varphi_{\mathrm{U2}}^0 &= 0.7635\state{04}-0.6458\state{11}, \\
    \varphi_{\mathrm{VX4}}^0 &= 0.7221\state{04}-0.6917i\state{11}.
\end{align}
These coefficients are estimated using the numerically exact states at $\mu=6.04$. The chemical potential and trap continuation of these states is summarized in Fig.~\ref{DS04a}.

The DS04 state has four horizontal dark soliton stripes in the vicinity of the linear limit. As the chemical potential increases, they connect into two dark soliton loops next to each other in the TF regime. It is quite striking that the evolution here is rather different from that of the DS04 state of $\kappa=1/2$. The DS04 state therein becomes a concentric deformed RDS2. The subsequent evolution in $\kappa$ therefore differs as well. As $\omega_x$ decreases, the two dark soliton loops move closer together, forming remarkably the DS8 state found earlier \cite{Wang:LLC}. To our knowledge, these two states are identical at $\kappa=1/2$.
As such, the two dark soliton loops also reconnect into a single loop at $\omega_x\approx1.89$, then it gradually morphs into the regular DS20 state in the isotropic trap as in \cite{Wang:LLC}.
%Maybe the DS04 state becomes the deformed RDS2? NO! The DS8 state has two neighbouring dark soliton loops facing each other, like the number $8$ or a sand glass.

The continuation of the DS11 state is robust in both the chemical potential and the trap, as expected. It becomes the XDS2 (cross dark soliton with two stripes) in the isotropic trap. The U2 soliton has two U solitons next to each other but open in opposite directions. The continuation in the chemical potential is quite robust. As it is continued into the isotropic trap, both U solitons gradually open an increasingly wider angle and then they connect into the DS11 state at $\omega_x \approx 2.43$.

The VX4 state is from a complex mixing of the \statem{04} and \statem{11} states in the weakly-interacting regime. It is an aligned vortex quadruple, and the vortex charge alternates except between the two central vortices. There are two small density depletion regions at the edge of the condensate. Interestingly, they are gradually healed as the chemical potential increases without any nucleation of vortices. The continuation in the chemical potential is quite robust, but the two central vortices approximately merge together in the TF regime.
As it is continued into the isotropic trap, the two central vortices essentially overlap from the $y$ direction but then they remarkably split again along the $x$ direction instead around $\omega_x=1.71$. It is quite intriguing that the final configuration in the isotropic trap is identical to the rhombus vortex quadruple state in \cite{Wang:LLC}. Four vortices of net charge $0$ have more isomers. In addition to the aligned vortex quadruple of alternating charge (see Eq.~\eqref{AVQ}), there is also a square vortex quadruple state \cite{Wang:LLC}.

%The SVQ is found to be quite robust except for a narrow interval of oscillatory instability \cite{PK:DSVX,Middelkamp:VX}.

%Note that the vortex charge alternates along both quadrilaterals. The RVQ is stable in a narrow chemical potential interval near the linear limit \cite{Panos:DC1}. 

The sixth linear degenerate set has two states \statem{05} and \statem{12} with $\mu_0=7$. 
The ULSs of the distinct wave patterns are:
\begin{align}
    \varphi_{\mathrm{DS05}}^0 &= \state{05}, \\
    \varphi_{\mathrm{DS12}}^0 &= \state{12}, \\
    \varphi_{\mathrm{U2I}}^0 &= 0.6865\state{05}-0.7271\state{12}, \\
    \varphi_{\mathrm{VX5}}^0 &= 0.7022\state{05}+0.7119i\state{12}.
\end{align}
These coefficients are estimated using the numerically exact states at $\mu=7.04$. The chemical potential and trap continuation of these waves is displayed in Fig.~\ref{DS04a}.

The DS05 has five horizontal dark soliton stripes in the vicinity of the linear limit. The continuation is qualitatively similar to that of the DS04 state, except there is an additional dark soliton stripe at $y=0$. As such, the two outer pairs connect into two dark soliton loops as the chemical potential increases. As it is continued into the isotropic trap, the two dark soliton loops again move closer together and reconnect into a single loop around $\omega_x=1.69$. Naturally, it morphs into the DS21 state in the isotropic trap \cite{Wang:LLC}. The continuation of the DS12 state is as expected, yielding a $\phi$ soliton in the TF regime, cf. the DS02 state above, which evolves into a RDS as shown in Fig.~\ref{GSa}.

The U2I soliton has two U solitons and a tilted dark soliton filament through the center in the near-linear regime. As the chemical potential increases, the three filaments connect into a single highly curved one. Nevertheless, the continuation in the chemical potential is pretty robust. As it is continued into the isotropic trap, the two U solitons gradually open a wider angle, and the dark soliton filaments connect into the $\phi$ soliton at $\omega_x\lesssim2.50$.

The VX5 is an aligned vortex quintopole state, the three central vortices have charge $1$ and the two side ones have charge $-1$ in the near-linear regime. As the chemical potential increases, each neighbouring vortices of the central one undergoes an elongation and pair creation at $\mu \approx 14.3$, yielding a VX9 state. In elongation and pair creation, a single vortex splits into three ones, the net charge is conserved and the charge of the central vortex is changed. Next, two far edge vortices of charge $-1$ are nucleated in our spatial horizon at $\mu \approx 15.3$, leading to a VX11 state in the TF regime. As $\omega_x$ decreases, the two vortices are induced into the condensate at $\omega_x\approx2.2$, and the central VX7 cluster also becomes substantially larger. It is remarkable that the state becomes the final chaotic VX9e state in the isotropic trap \cite{Wang:LLC}.

\begin{figure*}%{r}{0.5\textwidth}
\includegraphics[width=\textwidth]{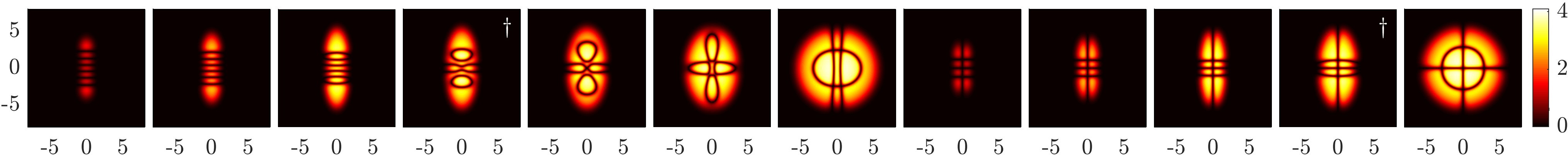}
\includegraphics[width=\textwidth]{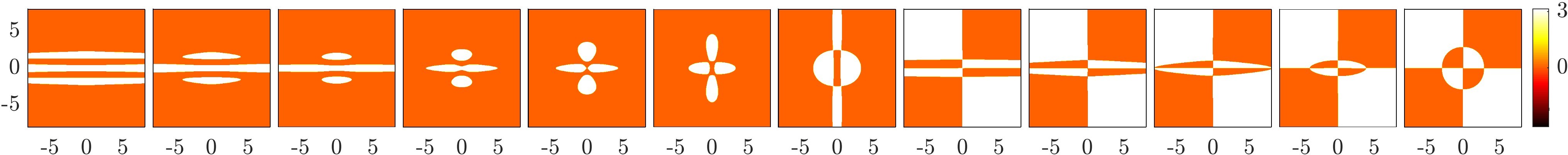}
\includegraphics[width=\textwidth]{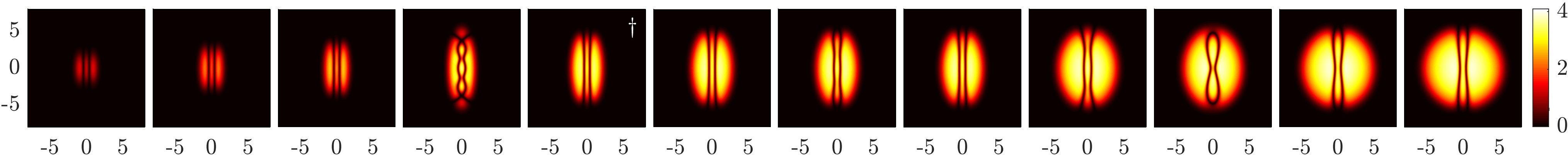}
\includegraphics[width=\textwidth]{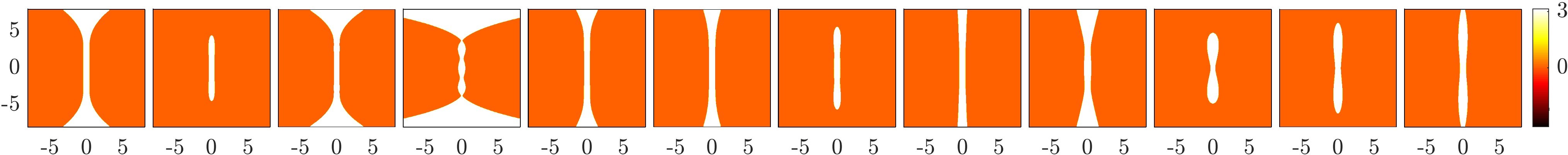}
\includegraphics[width=\textwidth]{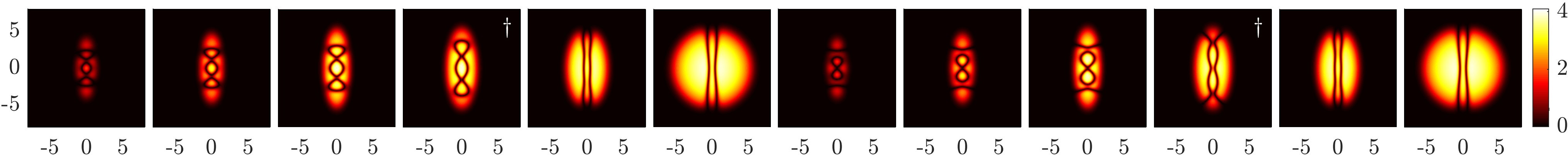}
\includegraphics[width=\textwidth]{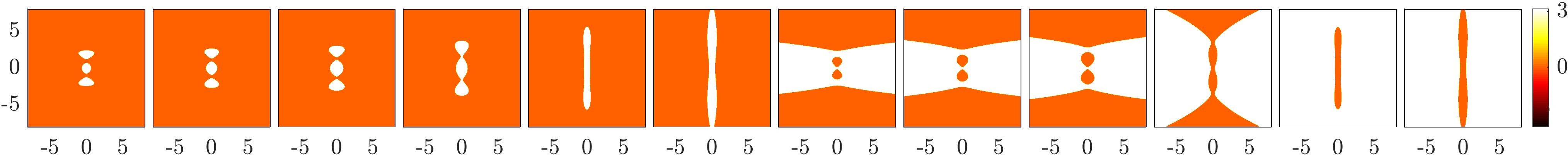}
\includegraphics[width=\textwidth]{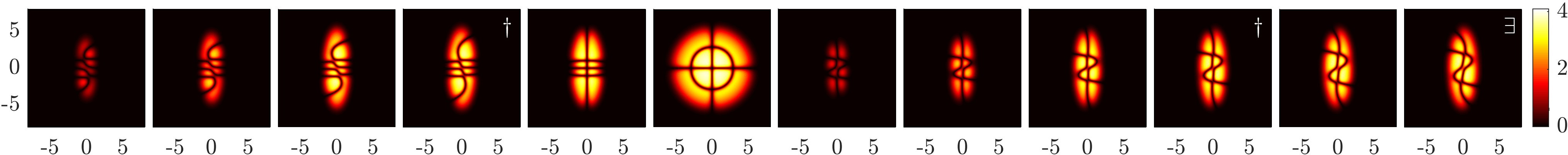}
\includegraphics[width=\textwidth]{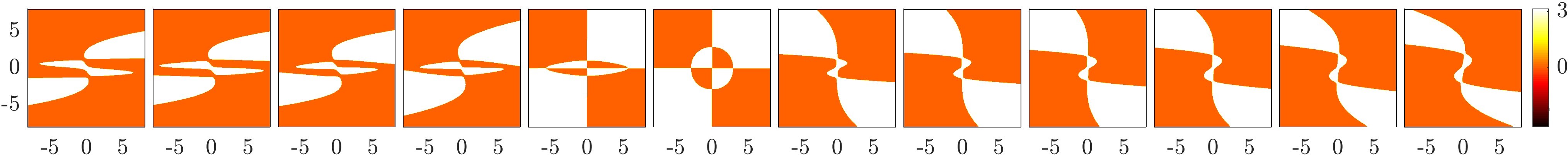}
\includegraphics[width=\textwidth]{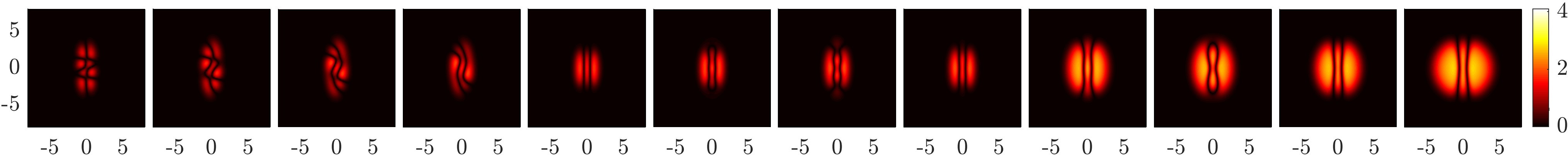}
\includegraphics[width=\textwidth]{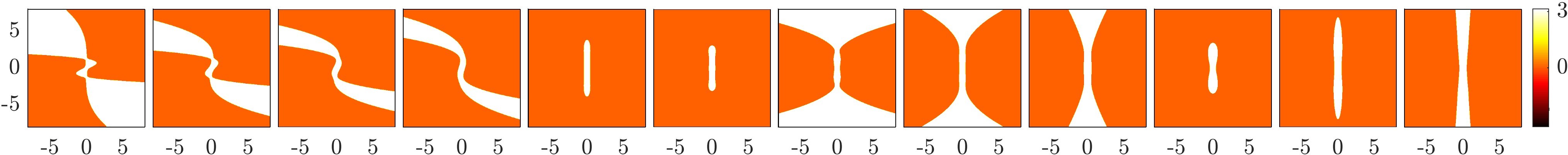}
\caption{
Solitary waves continued from the $\mu_0=8$ linear degenerate set of the $\kappa=1/3$ trap, part one. The first set shows the DS06 state at $\mu=9, 12, 16$ and $\omega_x=2.5, 2, 1.5, 1$, and the DS13 state at $\mu=9, 12, 16$ and $\omega_x=2, 1$. The second set depicts the DS20 state at $\mu=9, 11, 13, 16$ and $\omega_x=2.5, 1.92, 1.91, 1.88, 1.335, 1.333, 1.1, 1$. The third set illustrates the O3 and U2O2 states at $\mu=9, 12, 16$ and $\omega_x=2.6, 1.91, 1$, and $\omega_x=2.2, 1.91, 1$, respectively. 
The fourth set displays the U28 and W2 states at $\mu=9, 12, 16$ and $\omega_x=2.8, 2.5, 1$, and $\omega_x=2.9, 2.8, 2.757$, respectively. Here, the W2 state reaches an existence boundary highlighted by the exists symbol. However, the state can be continued into the isotropic trap in the near-linear regime, as shown in the final set at $\mu=9$. The states of $\omega_x=3, 2.9, 2.8, 2.7, 2.64, 2.5395, 2.5390, 2.45, 1.4526, 1.4524, 1.3, 1$ are sketched.
}
\label{DS06a}
\end{figure*}

\begin{figure*}[th]%{r}{0.5\textwidth}
\includegraphics[width=\textwidth]{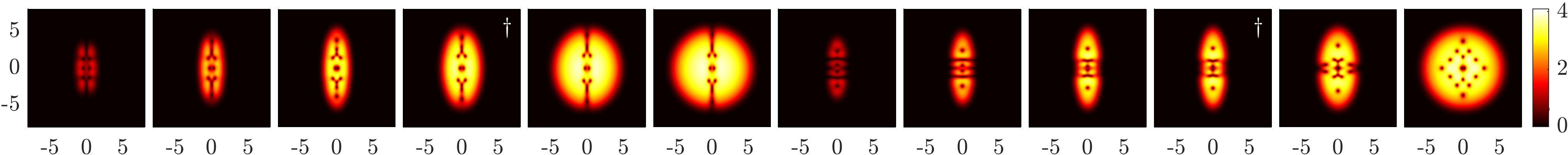}
\includegraphics[width=\textwidth]{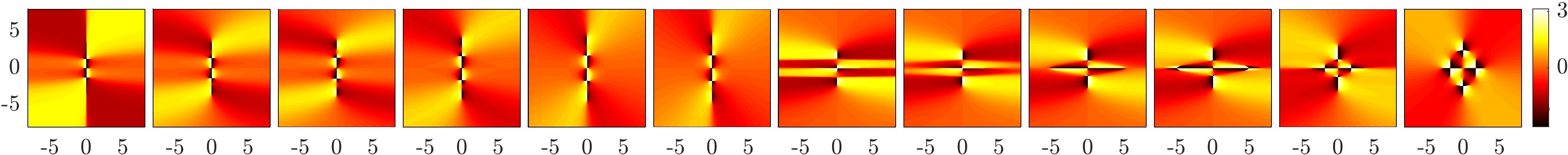}
\includegraphics[width=\textwidth]{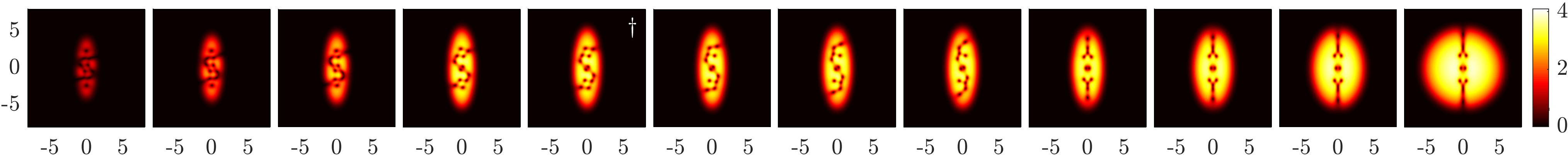}
\includegraphics[width=\textwidth]{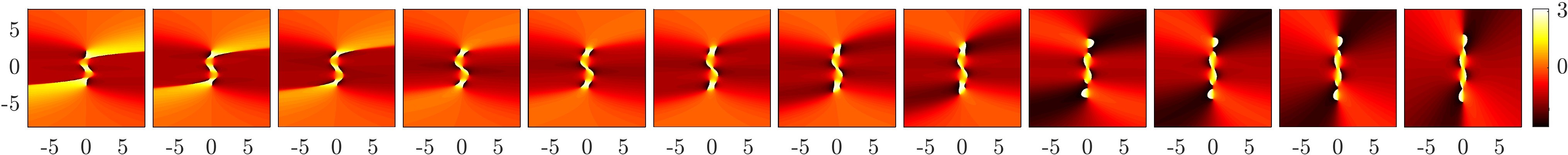}
\includegraphics[width=\textwidth]{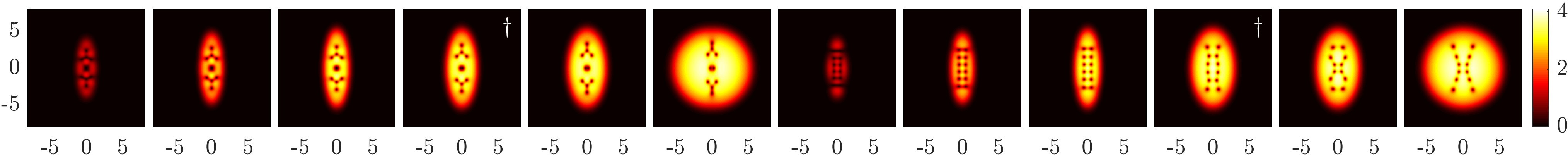}
\includegraphics[width=\textwidth]{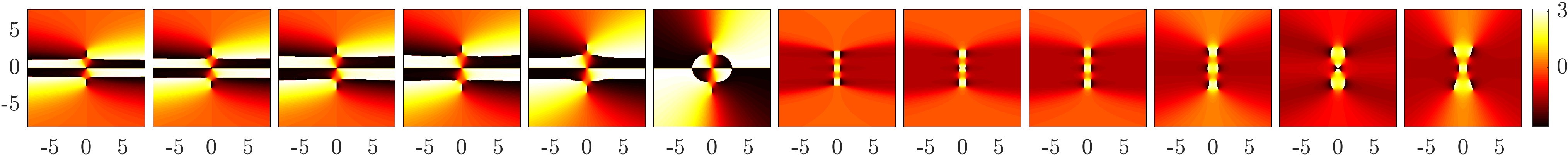}
\includegraphics[width=\textwidth]{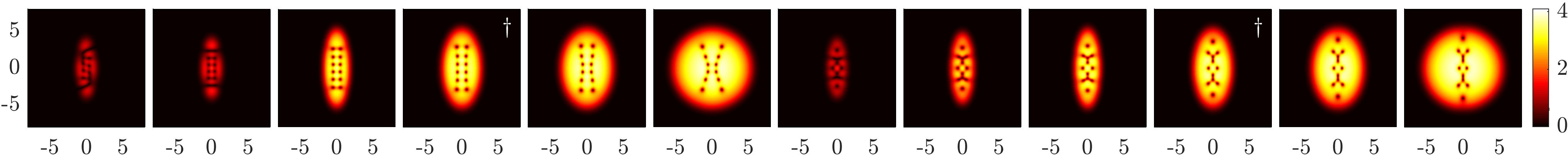}
\includegraphics[width=\textwidth]{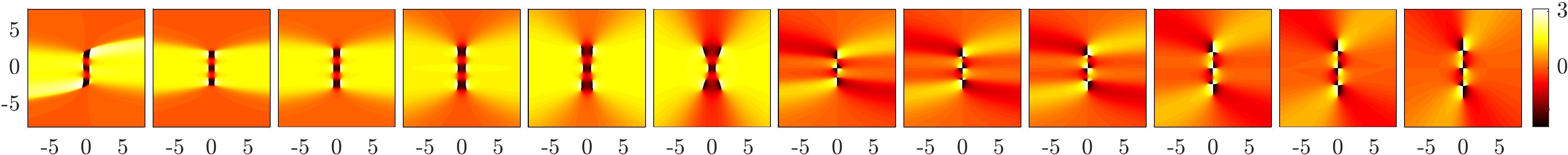}
\caption{
Solitary waves continued from the $\mu_0=8$ linear degenerate set of the $\kappa=1/3$ trap, part two. The first set shows the VX8a and VX8b states at $\mu=9, 13, 16$ and $\omega_x=2, 1.2, 1$, and $\omega_x=2.9, 2, 1$, respectively. The second set depicts the VX10a state at $\mu=9, 11, 13, 16$ and $\omega_x=2.9, 2.8, 2.7, 2.6, 2.5, 2, 1.5, 1$. The third set illustrates the VX10b and VX12a states at $\mu=9, 13, 16$ and $\omega_x=2.5, 2, 1$, and $\omega_x=1.77, 1.7657, 1$, respectively. The fourth set sketches the VX12b state at $\mu=9, 9.4, 16$ and $\omega_x=2, 1.5, 1$, and the VX12c state at $\mu=9, 13, 16$ and $\omega_x=2, 1.5, 1$.
}
\label{VX8a}
\end{figure*}

The seventh linear degenerate set has three states \statem{06}, \statem{13}, and \statem{20} with $\mu_0=8$. Finally, we reach the degeneracy of $3$ at the level of quantum number $6$, and the pattern formation becomes considerably richer, in line with the previous observation \cite{Wang:LLC}. The identified solitary waves as well as the chemical potential and trap continuations are shown in Fig.~\ref{DS06a} and Fig.~\ref{VX8a}.
Their respective ULSs are summarized here:
\begin{align}
    \varphi_{\mathrm{DS06}}^0 &= 0.99998\state{06}-0.0006\state{20}, \\
    \varphi_{\mathrm{DS13}}^0 &= \state{13}, \\
    \varphi_{\mathrm{DS20}}^0 &= -0.0077\state{06}+0.9999\state{20}, \\
    \varphi_{\mathrm{O3}}^0 &= 0.7982\state{06}+0.6024\state{20}, \\
    \varphi_{\mathrm{U2O2}}^0 &= 0.8052\state{06}-0.5930\state{20}, \\
    \varphi_{\mathrm{U28}}^0 &= 0.5959\state{06}-0.8006\state{13}+0.0627\state{20}, \\
    \varphi_{\mathrm{W2}}^0 &= 0.0665\state{06}+0.9480\state{13}+0.3112\state{20}, \\
    \varphi_{\mathrm{VX8a}}^0 &= 0.0049\state{06}+0.7914i\state{13}+0.6113\state{20}, \\
    \varphi_{\mathrm{VX8b}}^0 &= 0.6892\state{06}-0.7244i\state{13}+0.0134\state{20}, \\
    \varphi_{\mathrm{VX10a}}^0 &= 0.6938\state{06}+(0.0115+0.6094i)\state{13} \nonumber \\
    &-(0.0040+0.3836i)\state{20}, \\
    \varphi_{\mathrm{VX10b}}^0 &= 0.5908\state{06}+0.6803i\state{13}-0.4337\state{20}, \\
    \varphi_{\mathrm{VX12a}}^0 &= 0.7686\state{06}+(0.0098-0.6397i)\state{20}, \\
    \varphi_{\mathrm{VX12b}}^0 &= 0.6469\state{06}-(0.4445+0.0192i)\state{13} \nonumber \\
    &-(0.0173-0.6191i)\state{20}, \\
    \varphi_{\mathrm{VX12c}}^0 &= 0.5911\state{06}+0.6764i\state{13}+0.4393\state{20}.
\end{align}
These coefficients are estimated using the numerically exact states at $\mu=8.04$. 

The DS06 state essentially has a total of six horizontal dark soliton stripes in the vicinity of the linear limit. As the chemical potential increases, the top pair and the bottom pair connect into two dark soliton loops. As it is continued into the isotropic trap, the central pair also connects into a closed loop. Next, the central loop breaks and reconnects into two loops at $\omega_x \gtrsim 2.0$, leading to a four-petal structure. The structure then deforms further and it remarkably morphs into the double $\phi$ state in the isotropic trap \cite{Wang:LLC}.

The continuation of the DS13 state is qualitatively similar to that of the DS03 state in Fig.~\ref{GSa}, except that there is an additional vertical dark soliton stripe. As the chemical potential increases, the two outer horizontal dark soliton stripes tend to close. As it is continued into the isotropic trap, they connect into a single loop in our spatial horizon, and then gradually deform into the RDS, leading to the RDSXDS2 state in the isotropic trap.

The continuation of the DS20 state is essentially clear, but the details are somewhat complicated. As the chemical potential increases, the two dark soliton filaments connect into a single closed loop at $\mu \approx 9.2$, but then it opens again at $\mu \approx 12.4$. Next, the filaments significantly distort around $\mu=13.5$ and $\mu=14$, leading to a highly curved waveform in the TF regime. As it is continued into the isotropic trap, the filaments gradually become less distorted, e.g., they are quite smooth around $\omega_x=2.7$. The state of $\omega_x=2$ here is somewhat different from its counterpart of the $\kappa=1/2$ trap, the former is less curved than the latter \cite{Wang:LLC}. This is quite striking, as the DS20 is a quite low-lying state. This suggests that dark soliton filaments may take qualitatively similar but quantitatively different waveforms at a given vector parameter, leading to solitary wave isomers of dark soliton filaments. However, they become increasingly similar before they close into a single loop at $\omega_x \gtrsim 1.91$, and they become identical after this transition, to our knowledge. Interestingly, it quickly opens again at $\omega_x\approx1.89$. Next, the dark soliton filaments connect into a closed loop, making a transition to the DS8 state at $\omega_x\approx1.334$ \cite{Wang:LLC}. Finally, the state gradually morphs into the regular DS20 state of the isotropic trap. 

The O3 soliton has three aligned dark soliton loops, and its chemical potential continuation is robust. As it is continued into the isotropic trap, the dark soliton filaments reconnect into a single loop at $\omega_x\approx2.74$. We mentioned above that the DS20 state closes into a loop at $\omega_x \gtrsim 1.91$. The state after the transition is identical to the corresponding O3 state, to our knowledge. Therefore, the DS20 made a transition to the evolved O3 state, and consequently the subsequent evolution is identical. The loop opens around $\omega_x=1.89$, and it makes a transition to the DS8 state at $\omega_x\approx1.334$, and finally becomes the DS20 state of the isotropic trap.

The U2O2 soliton has two central dark soliton loops with a U soliton on each side along the $y$ axis, and its chemical potential continuation is robust. As it is continued into the isotropic trap, the two central loops reconnect at $\omega_x\approx2.85$, and then the full structure reconnects into the DS20 state at $\omega_x\approx2.43$. Interestingly, the state at $\omega_x=2$ is identical to the corresponding DS20 state of $\kappa=1/2$, therefore, these two states are parametrically connected \cite{Wang:LLC}. The dark soliton filaments remain quite curved at $\omega_x=2$, then it becomes quite smooth before they close into a loop at $\omega_x \gtrsim 1.91$. The following evolution is the same as that of the DS20 state above, therefore we should not discuss it again here for clarity. 

The U28 soliton has two U solitons with a deformed $8$ soliton in between, and the chemical potential continuation is robust. As it is continued into the isotropic trap, the outer branch of the two U solitons gradually opens and the dark soliton filaments quickly connect into the DS13 state above at $\omega_x \approx 2.57$. As such, the final state becomes the RDSXDS2 state in the isotropic trap.

The W2 soliton has two highly curved dark soliton filaments, but its chemical potential continuation is robust. As it is continued into the isotropic trap, the dark soliton filaments deform and it is striking that the state reaches an existence boundary at $\omega_x \approx 2.757$, to our knowledge. We continued the state further in the chemical potential, but the critical frequency only becomes slightly smaller. For example, the critical frequency is $\omega_x \approx 2.700$ at $\mu=24$. 
%However, it seems that the presence of an existence boundary is reasonable here, because the state is increasingly like an S soliton in coexistence with a vertical dark soliton in the TF regime. The S soliton tends to deform towards the vertical dark soliton, as mentioned earlier, see Fig.~\ref{GSa}. But this is impossible here as there is already a vertical dark soliton. It is unfeasible to have two dark solitons sitting on top of each other in a single component. 
It seems that dark soliton filaments are typically more flexible to continue compared with vortical structures because they can deform to a good extent, yet this example suggests that they may also encounter a nontrivial existence boundary. 

However, we have successfully continued the W2 soliton into the isotropic trap at $\mu=9$ in the near-linear regime. After a complicated series of structure deformations, it morphs into the DS20 state of the isotropic trap \cite{Wang:LLC}. %These two states become identical, to our knowledge. 
%This example shows again the complicated nature of solitary wave bifurcations. 
The two dark soliton filaments bend, smooth, and deform, and interestingly the state is like a double S soliton at $\omega_x \approx 2.8$; cf. the S soliton in Fig.~\ref{GSa}. The double S soliton bends towards the vertical direction and it becomes a single closed loop at $\omega_x \approx 2.64$. The following evolution is quite similar to that of the DS20 state above. The dark soliton loop deforms and then it suddenly opens at $\omega_x \approx 2.54$. The dark soliton filaments smooth quickly, and then they close into the $8$ soliton at $\omega_x \approx 1.45$. Finally, the DS8 gradually opens again and turns into the regular DS20 state in the isotropic trap.

Next, we turn to vortical patterns, as illustrated in Fig.~\ref{VX8a}. The VX8a state has eight vortices, six of them are inside the condensate and two of them are edge vortices, and the vortex charge alternates along the loop. As the chemical potential increases, two pair creations occur in the density depletion region at $\mu \lesssim 13.0$, and also the two edge vortices are induced into the condensate at $\mu \approx 13$, forming a VX12 structure. The Y clusters are quite common in vortical patterns \cite{Wang:LLC}. The two central vortices of the same charge are also pushed closer together. Interestingly, the process is roughly undone as the state is continued into the isotropic trap. The two vortices again become edge vortices at $\omega_x\approx1.3$, and then two pair annihilations take place at $\omega_x\approx1.17$, yielding again the VX8a state. Finally, the two central vortices are also pushed further apart. %This state is a new state obtained from the $\kappa=1/3$ trap, it was not found in \cite{Wang:LLC}.

The VX8b state contains eight vortices, six of them are along the $y$ axis and two of them are along the $x$ axis. The inner four vortices of the former are negatively charged, and the rest of them are positively charged. As the chemical potential increases, each of the second and the fifth vortices along the $y$ axis undergoes an elongation and pair creation at $\mu \approx 14.0$, leading to a VX12 state. As such, the neighbouring vortices have the opposite charges. As $\omega_x$ decreases, each of the edge vortices also rapidly undergoes an elongation and pair creation at $\omega_x \approx 2.97$, yielding a VX16 state. Next, the two far edge vortices are temporally ejected out at $\omega_x \approx 2.8$ but they nucleate again at $\omega_x \approx 2.2$ in our spatial horizon. Note that they remain far away from the bulk of the condensate in this process. The four inner vortices and the two outer vortices are induced into the condensate at $\omega_x \approx 2.2$ and $\omega_x \approx 1.8$, respectively. Interestingly, the central two vortices merge along the $y$ axis and then they split again along the $x$ axis at $\omega_x \approx 1.66$. It is remarkable that the state in the isotropic trap becomes the final chaotic VX12d state of $\kappa=1$ \cite{Wang:LLC}.

The VX10a state has a complicated vortical waveform, and it is largely from a complex mixing of the W2 soliton and the DS06 state. As the chemical potential increases, two far edge vortices are quickly induced at $\mu \gtrsim 8.1$ in our spatial horizon, yielding a VX12 state. The subsequent continuation is quite robust, the wave pattern slightly deforms and the vortices become more prominent as the density grows, and the two edge vortices are also induced into the condensate at $\mu \approx 15$. As it is continued into the isotropic trap, the two vortices gradually rotate towards the $y$ axis, the other vortices also gradually reorganize, and then the state remarkably becomes symmetric and converges to the VX8a state above at $\omega_x \gtrsim 2.52$. Therefore, the subsequent evolution is identical, and we should not discuss it further here.

The VX10b state has two Y clusters of central vortex charge $1$ on each side, and there are two central vortices of charge $1$ along the $y$ axis. The chemical potential continuation is robust, and the two central vortices are pushed closer together in the TF regime. As it is continued into the isotropic trap, the two central vortices quickly merge along the $y$ axis and then they split again along the $x$ axis at $\omega_x \approx 2.94$. The subsequent continuation is very robust.

The VX12a state is like a miniature $6\times 2$ vortex lattice, as it is largely from a complex mixing of the \statem{06} and \statem{20} states. The neighbouring vortices have the opposite charge, and the continuation of the state is robust in the chemical potential. As it is continued into the isotropic trap, the lattice becomes slightly distorted until it makes a discontinuous state transition between about $\omega_x=1.77$ and $\omega_x=1.7657$. The two vortical stripes morph into a single loop structure like the $8$ dark soliton. It seems likely that this transition is linked to the corresponding transition of the DS20 state above in Fig.~\ref{DS06a}. Next, the two stripes gradually emerge again as they become more straight at $\omega_x \approx 1.6$. The two ends are slightly wider, and they become increasingly so as the state is continued into the isotropic trap. 
%We emphasize that the two states of the discontinuous transition are weakly parametrically connected in the sense that there is a series of quasi-states linking them in the Newton's convergence process \cite{Wang:LLC}. We present the states here for completeness to find more solitary waves, and readers who are not willing to establish such a parametric connectivity may ignore the latter part, then there is an existence boundary.

The VX12b state is similar in structure to the VX12a state, except that the lattice is somewhat distorted. The continuation in the near-linear regime is pretty robust, until it makes a transition and becomes symmetric at $\mu \approx 9.39$. This final state is essentially identical to the VX12a state above, to our knowledge. Therefore, the subsequent evolution in both the chemical potential and the trap frequency is the same, and we should not discuss it again here for simplicity.

The VX12c state has $4$ negatively charged vortices in a rhombus shape, with $2$ additional Y clusters of negatively charged central vortices. Clearly, the outer vortices of the Y clusters are oppositely charged compared with the central ones. The continuation in both the chemical potential and the trap is very robust.

\subsection{Wave patterns from trap aspect ratio 2/3}

The $\kappa=2/3$ lattice planes strike through a number of nondegenerate low-lying linear states. They do not lead to new solitary waves here, as the pertinent waves have already been found in other traps. Nevertheless, the ULSs are:
\begin{align}
    \varphi_{\mathrm{GS}}^0 &= \state{00}, \quad \mu_0=1.25, \\
    \varphi_{\mathrm{DS01}}^0 &= \state{01}, \quad \mu_0=2.25, \\
    \varphi_{\mathrm{DS10}}^0 &= \state{10}, \quad \mu_0=2.75, \\
    \varphi_{\mathrm{DS02}}^0 &= \state{02}, \quad \mu_0=3.25, \\
    \varphi_{\mathrm{DS11}}^0 &= \state{11}, \quad \mu_0=3.75.
\end{align}
The continuation of these states is robust, and careful inspection shows that these states are parametrically connected with the corresponding states of the anisotropic traps of $\kappa=1/2$ and $\kappa=1/3$. They yield the ground state, the horizontal dark soliton stripe, the vertical dark soliton stripe, the RDS, and the XDS2 state in the isotropic trap, respectively. As their continuation is essentially similar to that of their counterparts, we should not discuss it further here for simplicity.

%The ground states are all parametrically connected. We have checked the four final states in the isotropic trap, and the other cases as well.

%The dark soliton stripe states are all parametrically connected. We have checked the seven final states of DS10 or DS01 of the four traps. In addition, the final state of the U dark soliton of $\kappa=1/2$ and the S dark soliton of $\kappa=1/3$ are also compared. Then, the other cases of the anisotropic traps are compared as well.

\begin{figure*}[t]
\includegraphics[width=\textwidth]{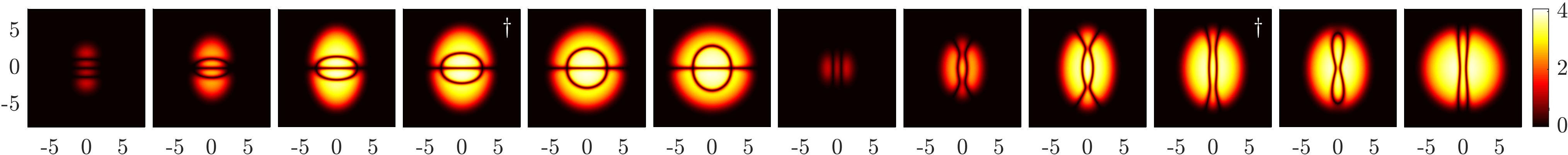}
\includegraphics[width=\textwidth]{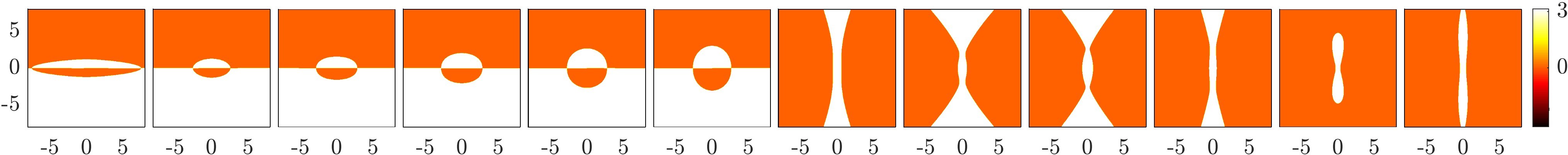}
\includegraphics[width=\textwidth]{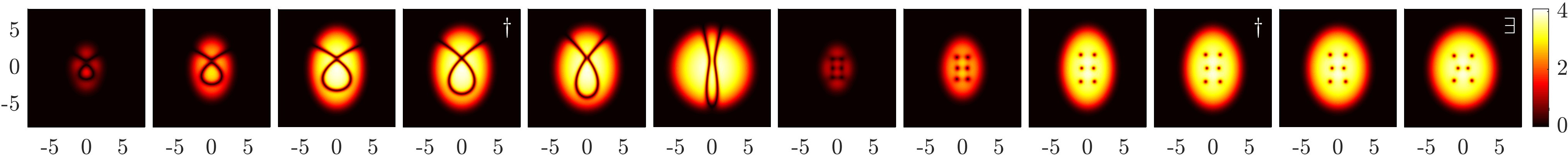}
\includegraphics[width=\textwidth]{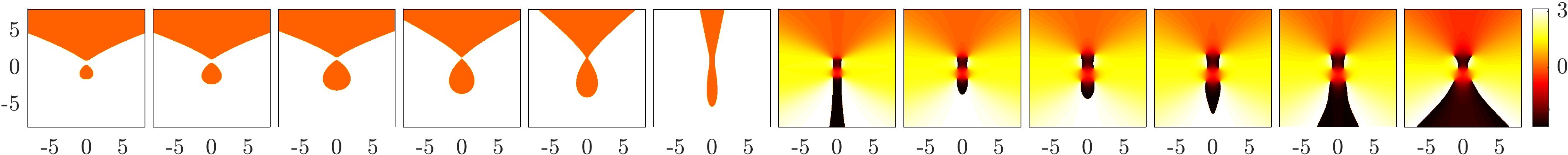}
\includegraphics[width=\textwidth]{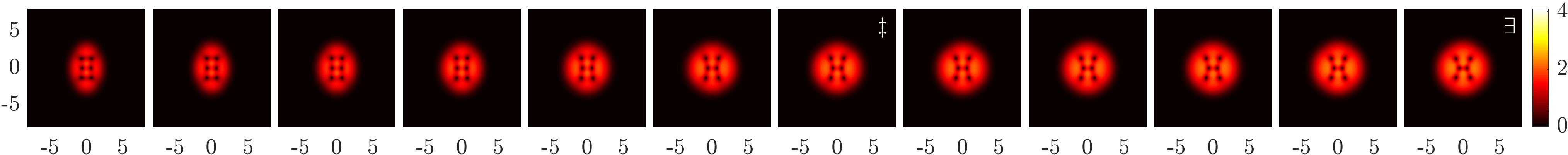}
\includegraphics[width=\textwidth]{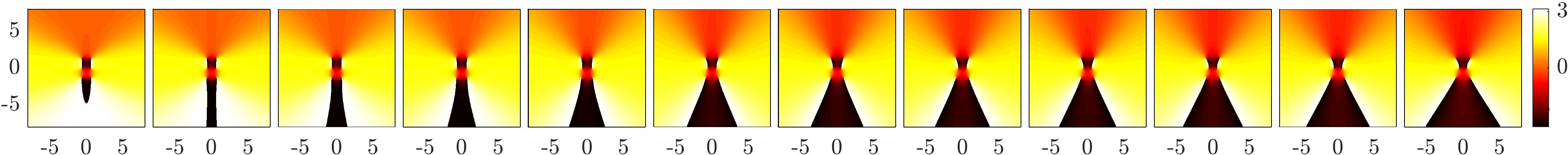}
\caption{
Solitary waves continued from the $\mu_0=4.25$ linear degenerate set of the $\kappa=2/3$ trap. The first set shows the DS03 and DS20 states at $\mu=5, 9, 16$ and $\omega_x=1.3, 1.1, 1$, and $\omega_x=1.335, 1.333, 1$, respectively. The second set depicts the UO soliton at $\mu=5, 9, 16$ and $\omega_x=1.4, 1.3, 1$, and the VX6 state at $\mu=5, 9, 16$ and $\omega_x=1.4, 1.3, 1.2104$. Here, the VX6 state reaches an existence boundary. However, the state can be continued into the isotropic trap in the near-linear regime, as shown in the final set at $\mu=6$. The states of $\omega_x=1.5, 1.4, 1.3, 1.2, 1.1, 1$ are illustrated. Finally, this state is continued further in the chemical potential, as highlighted by the double dagger symbol. Here, the states of $\mu=6.1, 6.2, 6.3, 6.4, 6.5, 6.66$ are sketched, reaching again an existence boundary.
}
\label{DS03Aa}
\end{figure*}

The sixth linear degenerate set has two states \statem{03} and \statem{20} with eigenenergy $\mu_0=4.25$. It is from here that we start to harvest new solitary waves. We have identified a total of four states, and their respective ULSs are:
\begin{align}
    \varphi_{\mathrm{DS03}}^0 &= \state{03}, \\
    \varphi_{\mathrm{DS20}}^0 &= \state{20}, \\
    \varphi_{\mathrm{UO}}^0 &= 0.7509\state{03}-0.6604\state{20}, \\
    \varphi_{\mathrm{VX6}}^0 &= 0.7300\state{03}+0.6834i\state{20}.
\end{align}
These coefficients are estimated using the numerically exact states at $\mu=4.29$. The chemical potential and trap continuation of these states is shown in Fig.~\ref{DS03Aa}.

We have already continued the DS03 state in the $\kappa=1/2$ and $\kappa=1/3$ traps, and this state here is parametrically connected with the earlier counterparts. As such, the dark soltion filaments connect into the $\phi$ soliton as the chemical potential increases, and then it becomes the regular $\phi$ soliton in the isotropic trap. The continuation is robust as expected. As mentioned earlier, the $\phi$ soliton has its own linear limit in the isotropic trap \cite{Wang:LLC}.

The continuation of the DS20 state in the chemical potential is robust, however, it is interesting that the DS20 state in the TF regime differs from the $\kappa=1/2$ and $\kappa=1/3$ counterpart at $\omega_x=1.5$. The former is curved and further separated, while the latter is almost straight and closer together. As it is continued into the isotropic trap, the states gradually become essentially the same, and then the state also undergoes the transition to the DS8 state at $\omega_x\approx1.334$. These final states after the transition are identical, to our knowledge. As such, the state subsequently morphs into the regular DS20 state of the isotropic trap. The overall evolution of the DS20 states from various traps shows that the parametric connectivity of similar looking solitary waves can be rather intriguing and complicated.

The UO soliton has a U dark soliton and a ring dark soliton along the $y$ axis. It is a parity symmetry-breaking state, note that the two basis states have opposite parities. By contrast, the pertinent basis states of each linear degenerate set of $\kappa=1/3$ have a definite parity. The chemical potential continuation is robust, and the two dark soliton filaments move closer in the TF regime. As $\omega_x$ decreases, the filaments undergo a dark soliton reconnection at $\omega_x\approx1.39$, forming a single dark soliton filament. It is interesting that the state remains asymmetric in the isotropic trap, demonstrating that asymmetric states may persist into the isotropic trap. This is essentially the first asymmetric state we have continued into the isotropic trap and it remains asymmetric. It is reasonably common to find asymmetric states in anisotropic traps, but many of them become quite symmetric when they are continued into the isotropic trap; cf. the U soliton, the $\Psi$ soliton, and the U2 soliton of the $\kappa=1/2$ trap \cite{Wang:LLC}.

The VX6 state has two aligned VX3 stripes of alternating charge from the complex mixing of the \statem{03} and \statem{20} states. While the chemical potential continuation is quite robust, it seems that the trap continuation is challenging and the state survivals only upto $\omega_x \gtrsim 1.21$. The critical frequency $\omega_x$ only decreases slightly as the chemical potential decreases in the TF regime. The two stripes bend as $\omega_x$ decreases, and they also tend to bend as the chemical potential increases. This explains the observed trend between the critical frequency and the chemical potential. However, the VX6 state is successfully continued into the isotropic trap at $\mu=6$ in the near-linear regime. Finally, we continue this state further in the chemical potential, finding an upper critical chemical potential at $\mu_c \approx 6.66$. This state was found by the two-mode analysis \cite{Middelkamp:VX} and also the deflation method, featuring a narrow stable interval of $\mu \in [4.3, 4.4]$ in the isotropic trap \cite{Panos:DC1}. 

Next, we find another isolated basis state:
\begin{align}
    \varphi_{\mathrm{DS12}}^0 &= \state{12}, \quad \mu_0=4.75.
\end{align}
This appears to be the final isolated state in this set of lattice planes, and naturally it leads to the $\phi$ soliton in the TF regime and the isotropic trap. The continuation is very similar to its counterparts of $\kappa=1/2$ and $\kappa=1/3$, and these states are parametrically connected, to our knowledge. Therefore, we should not discuss it further here.

\begin{figure*}[t]
\includegraphics[width=\textwidth]{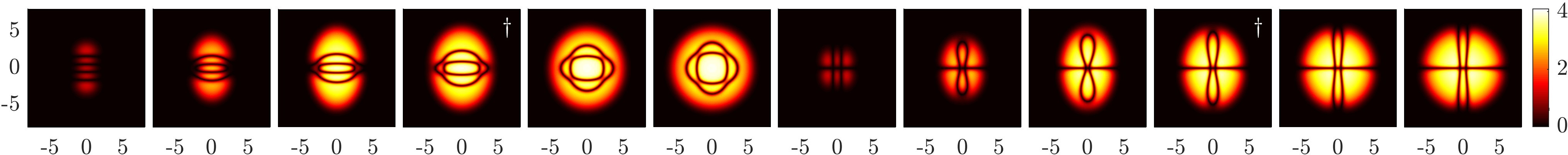}
\includegraphics[width=\textwidth]{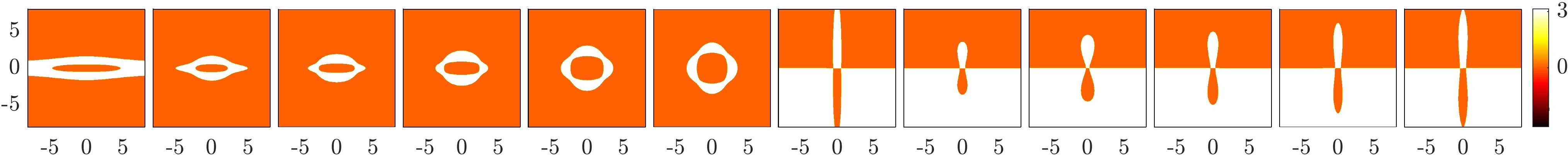}
\includegraphics[width=\textwidth]{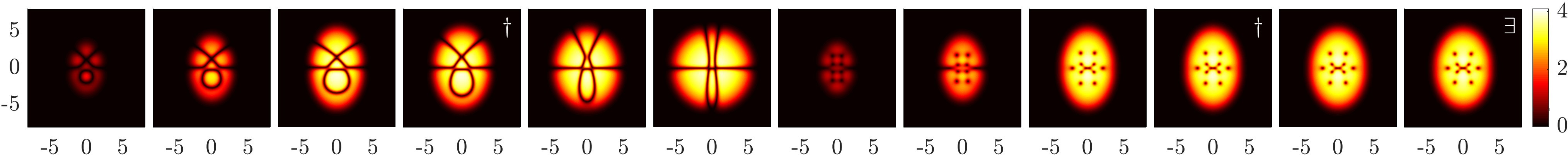}
\includegraphics[width=\textwidth]{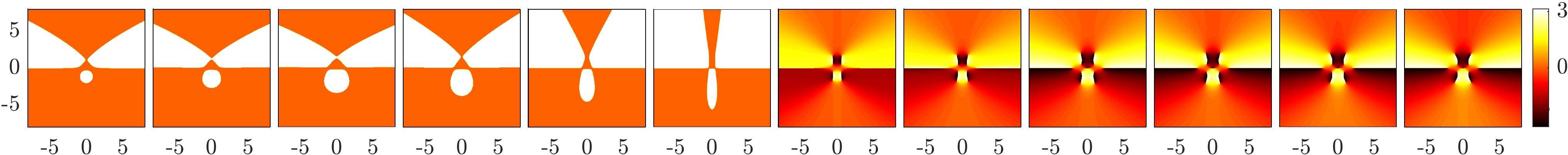}
\includegraphics[width=\textwidth]{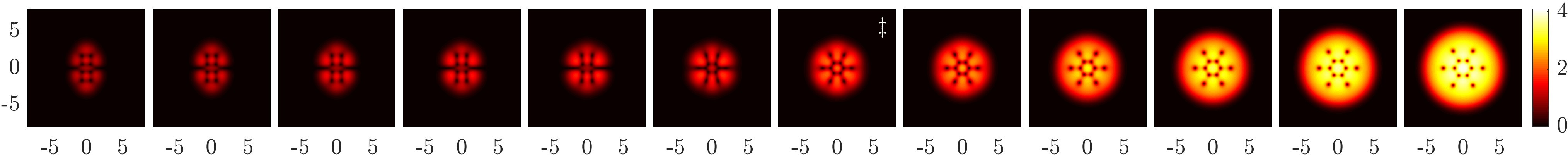}
\includegraphics[width=\textwidth]{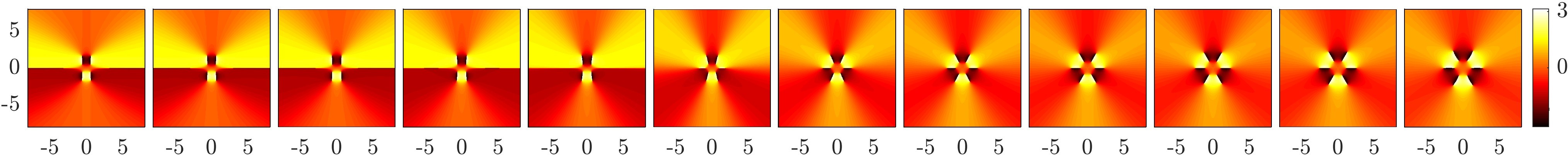}
\caption{
Solitary waves continued from the $\mu_0=5.25$ linear degenerate set of the $\kappa=2/3$ trap. The first set shows the DS04 and DS21 states at $\mu=6, 10, 16$ and $\omega_x=1.3, 1.1, 1$. The second set depicts the U2O soliton at $\mu=6, 10, 16$ and $\omega_x=1.4, 1.2, 1$, and the VX8 state at $\mu=6, 10, 16$ and $\omega_x=1.45, 1.4, 1.3890$, reaching an existence boundary. However, this state can be continued into the isotropic trap at $\mu=6$ in the near-linear regime, as shown in the final set, and the states of $\omega_x=1.5, 1.4, 1.3, 1.2, 1.1, 1$ are illustrated. Finally, this state in the isotropic trap is continued further in the chemical potential, and the states of $\mu=7, 8, 10, 12, 14, 16$ are sketched.
}
\label{DS04Aa}
\end{figure*}

The eighth linear degenerate set has two basis states \statem{04} and \statem{21} with eigenenergy $\mu_0=5.25$. We have identified a total of four states, and their respective ULSs are:
\begin{align}
    \varphi_{\mathrm{DS04}}^0 &= \state{04}, \\
    \varphi_{\mathrm{DS21}}^0 &= \state{21}, \\
    \varphi_{\mathrm{U2O}}^0 &= 0.6607\state{04}-0.7506\state{21}, \\
    \varphi_{\mathrm{VX8}}^0 &= 0.6868\state{04}+0.7268i\state{21}.
\end{align}
These coefficients are estimated using the numerically exact states at $\mu=5.29$. The continuation of these states is illustrated in Fig.~\ref{DS04Aa}.

The DS04 state has four horizontal dark soliton stripes in the weakly-interacting regime. As the chemical potential increases, it forms two loops with an inner one and an outer one, similar to its counterpart of $\kappa=1/2$ \cite{Wang:LLC}. Indeed, the two states are parametrically connected, to our knowledge. By contrast, the counterpart of $\kappa=1/3$ forms two loops with an upper one and a lower one, as shown in Fig.~\ref{DS04a}. This is presumably because the condensate therein is more elongated. As $\omega_x$ decreases, the two loops gradually deform, yielding two rounded square loops. The inner loop is rotated with respect to the outer one by $\pi/4$. This deformed RDS2 state is an isomer of the symmetric RDS2 polar state \cite{Wang:LLC}. 

The continuation of the DS21 state is quite robust, contrary to the above DS20 state in Fig.~\ref{DS03Aa}. It is striking that an additional horizontal dark soliton can have such a strong effect on the continuation. Therefore, the final state is parametrically connected to the regular DS21 state of the isotropic trap, to our knowledge.

The U2O soliton has two U solitons and an O soliton in the low-density regime. The two U dark soliton filaments undergo a dark soliton reconnection at $\mu \approx 8.3$. As it is continued into the isotropic trap, the two upper dark soliton filaments reconnect back to the original configuration qualitatively at $\omega_x\approx1.44$. Next, the filaments deform further and the final state in the isotropic trap is like the evolved UO state in Fig.~\ref{DS03Aa} with an additional horizontal dark soliton. As such, this is also a parity symmetry-breaking state that persists into the isotropic trap, see its phase profile.

The VX8 state has two aligned VX4 stripes, the vortex charge alternates between the two stripes, but the central vortices have the same charge vertically in each stripe. As the chemical potential increases, two edge vortices are quickly induced into the two density depletion wedges, yielding a VX10 state. The two edge vortices are induced into the condensate at $\mu \approx 12$. Next, a pair creation occurs in the center of the condensate at $\mu \approx 14.1$ such that the vortex change alternates between the nearest neighbours. The trap continuation is regular until encountering an existence boundary at $\omega_x \approx 1.3890$. The critical frequency diminishes only slightly as the chemical potential decreases, however, the continuation is successfully at $\mu=6$ in the near-linear regime. As the VX10 (the evolved VX8) state of $\mu=6$ is continued into the isotropic trap, two additional edge vortices are induced into the two density depletion wedges at $\omega_x \approx 1.22$ in our spatial horizon, leading to a VX12 state. Next, two pair annihilations quickly occur at $\omega_x \approx 1.19$ for the four induced vortices, yielding a VX8 state. Subsequently, two pair creations take place therein at $\omega_x \approx 1.03$ and therefore the VX8 state turns back into a VX12 state. Finally, we continue this state further in the chemical potential in the isotropic trap. Interestingly, it becomes a double vortex necklace of size $6$ state. This state is also obtained indirectly from a completely different path at a larger chemical potential $\mu=20$ \cite{Wang:LLC}. In this vein, we continue the state here further to $\mu=20$ and confirm that they are identical, to our knowledge. Two concentric vortex necklaces of sizes $4$ and $8$, respectively, are obtained in \cite{Wang:LLC}. Particularly, the latter state has a linear limit in the isotropic trap. A similar state of size $10$ is obtained in a rotating condensate \cite{RCG:rotating}, showing that there is likely a series of such states following this pattern.

%The VX16b state has two concentric vortex necklaces, and each ring has eight vortices. The vortex charge alternates along each ring, and also between the neighbouring vortices of the two rings. Both states are subject to oscillatory instabilities in the vicinity of the linear limit \cite{Panos:DC1}.

\begin{figure*}[t]
\includegraphics[width=\textwidth]{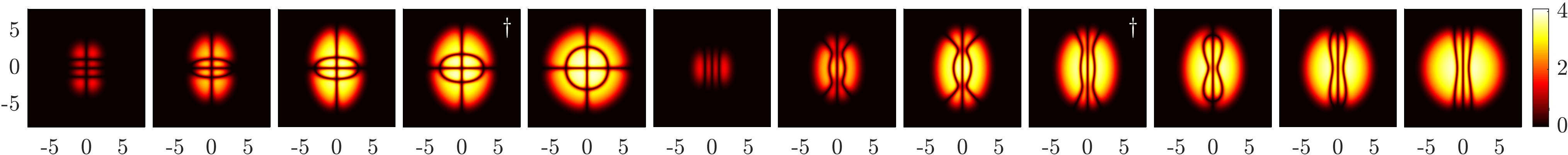}
\includegraphics[width=\textwidth]{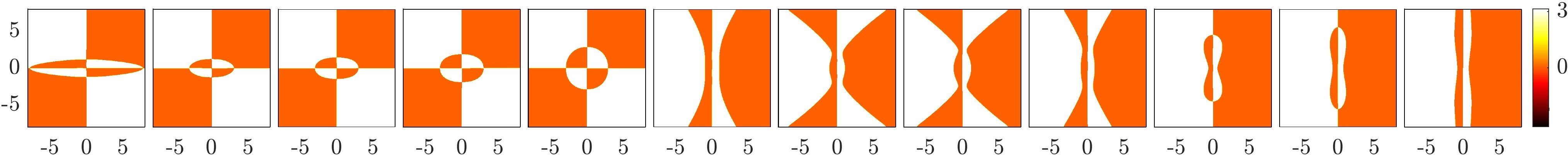}
\includegraphics[width=\textwidth]{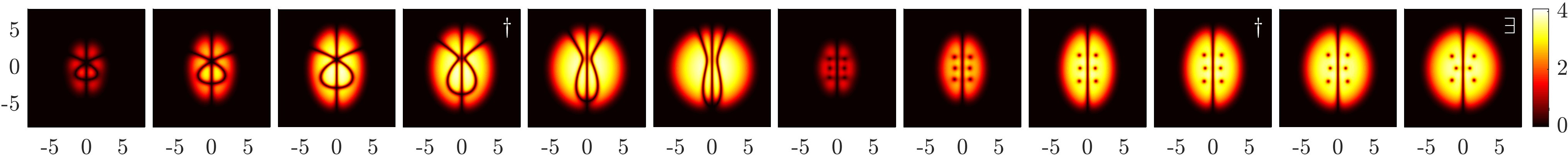}
\includegraphics[width=\textwidth]{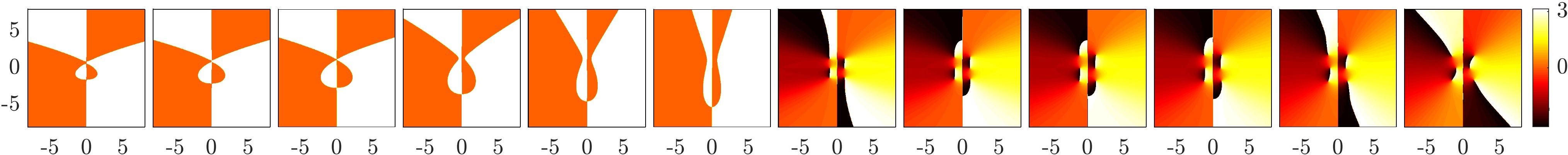}
\includegraphics[width=\textwidth]{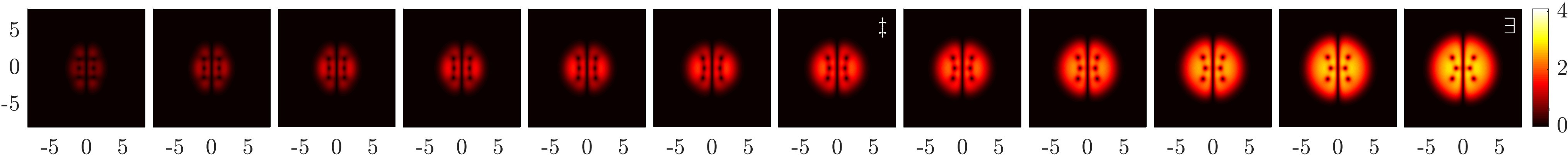}
\includegraphics[width=\textwidth]{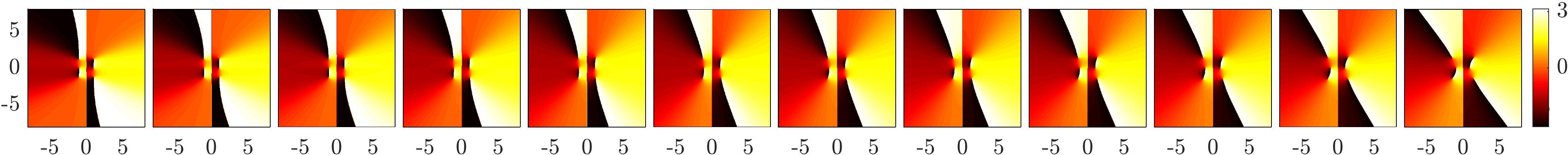}
\caption{
Solitary waves continued from the $\mu_0=5.75$ linear degenerate set of the $\kappa=2/3$ trap. The first set shows the DS13 state at $\mu=7, 11, 16$ and $\omega_x=1.3, 1$, and the DS30 state at $\mu=7, 11, 16$ and $\omega_x=1.37, 1.36, 1.2, 1$. The second set depicts the $\Psi\mathrm{O}$ soliton at $\mu=7, 11, 16$ and $\omega_x=1.3, 1.1, 1$, and the DSVX6 state at $\mu=7, 11, 16$ and $\omega_x=1.4, 1.2, 1.1095$, reaching an existence boundary. However, this state can be continued into the isotropic trap in the near-linear regime, as shown in the final set at $\mu=6$. The states of $\omega_x=1.5, 1.4, 1.3, 1.2, 1.1, 1$ are illustrated. Finally, the state in the isotropic trap is continued further in the chemical potential, and the states of $\mu=6.5, 7, 8, 9, 10, 10.495$ are sketched, reaching again an existence boundary.
}
\label{DS13Aa}
\end{figure*}

The ninth linear degenerate set has two basis states \statem{13} and \statem{30} with eigenenergy $\mu_0=5.75$. We have identified a total of four states, and their respective ULSs are:
\begin{align}
    \varphi_{\mathrm{DS13}}^0 &= \state{13}, \\
    \varphi_{\mathrm{DS30}}^0 &= \state{30}, \\
    \varphi_{\Psi\mathrm{O}}^0 &= 0.8307\state{13}-0.5566\state{30}, \\
    \varphi_{\mathrm{DSVX6}}^0 &= 0.7660\state{13}+0.6428i\state{30}.
\end{align}
These coefficients are estimated using the numerically exact states at $\mu=5.79$. The continuation of these states is illustrated in Fig.~\ref{DS13Aa}.

The continuation of the DS13 state is quite robust. First, the two outer dark soliton stripes along the $y$ axis quickly connect, forming a loop structure. As $\omega_x$ decreases, it gradually deforms into the RDSXDS2 state in the isotropic trap. The continuation is very similar to its counterpart of $\kappa=1/3$ in Fig.~\ref{DS06a}. Indeed, these two states are parametrically connected.

The chemical potential continuation of the DS30 state is quite robust, despite that it becomes quite distorted in the TF regime. As it is continued into the isotropic trap, the DS30 undergoes a discontinuous transition at $\omega_x\approx1.365$, and the two outer dark soliton stripes form a DS8 structure. Note that we have observed this kind of transition several times, see, e.g., the DS20 state in Fig.~\ref{DS03Aa}. Finally, the dark soliton loop gradually opens again in our spatial horizon and the final configuration becomes the regular DS30 state in the isotropic trap.

The $\Psi\mathrm{O}$ state has a $\Psi$ shape in the vicinity of the linear limit, however, it quickly undergoes a dark soliton reconnection at $\mu \approx 5.79$ and consequently a dark soliton loop emerges. The chemical potential continuation is pretty robust. As $\omega_x$ decreases, it is interesting that the state quickly undergoes another dark soliton reconnection at $\omega_x \approx 1.49$ and restores the $\Psi$ structure. The final state slightly differs from the regular DS30 state, and it is a parity symmetry-breaking state, see the phase profile. It seems likely that the final UO state of Fig.~\ref{DS03Aa} is a basic structure, and the final U2O state of Fig.~\ref{DS04Aa} possesses an additional horizontal dark soliton stripe while the final state here features an additional vertical one.

The DSVX6 state has an array of $3\times2$ vortices of alternating charge, in coexistence with a dark soliton stripe along the $y$ axis. While the chemical potential continuation is quite robust, the state reaches an existence boundary at $\omega_x \approx 1.1095$. This critical frequency only diminishes very slowly with decreasing chemical potential in the TF regime. However, the state is successfully continued into the isotropic trap at $\mu=6$ in the near-linear regime. Finally, we further continue the state in chemical potential in the isotropic trap, until it reaches an existence boundary at $\mu_c\approx10.50$. The DSVX6 state is spectrally unstable in the isotropic trap \cite{Middelkamp:VX}.

%The Jump2 breaks this limit by a more accurate estimator. However, it is possible that this is an illustration of numerical accuracy. We are checking if it also reaches a similar limit, or the state continues at higher mu. It seems that this is true, there is a new existence boundary at $\mu\approx10.501(1)$.

\begin{figure*}[t]
\includegraphics[width=\textwidth]{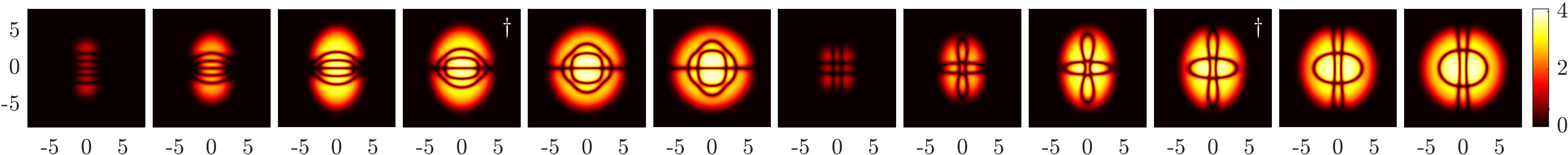}
\includegraphics[width=\textwidth]{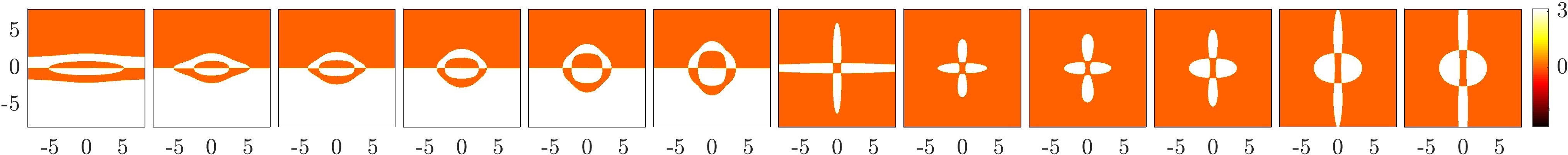}
\includegraphics[width=\textwidth]{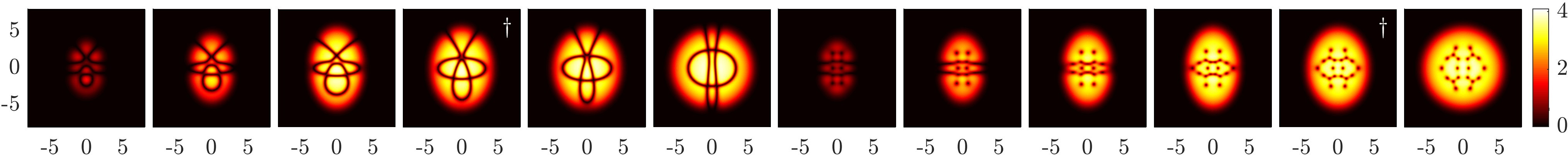}
\includegraphics[width=\textwidth]{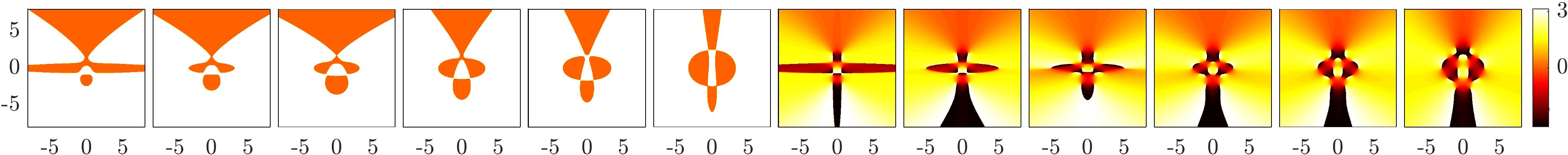}
\includegraphics[width=\textwidth]{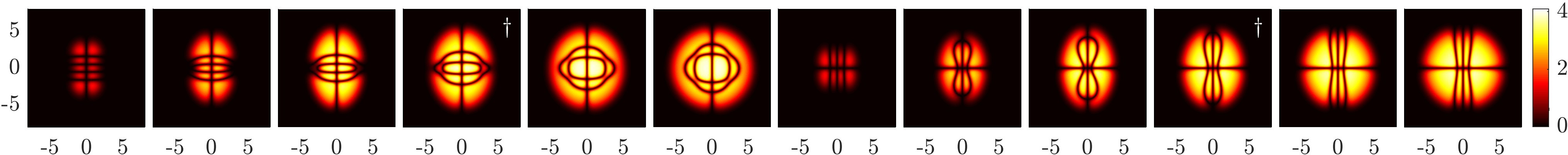}
\includegraphics[width=\textwidth]{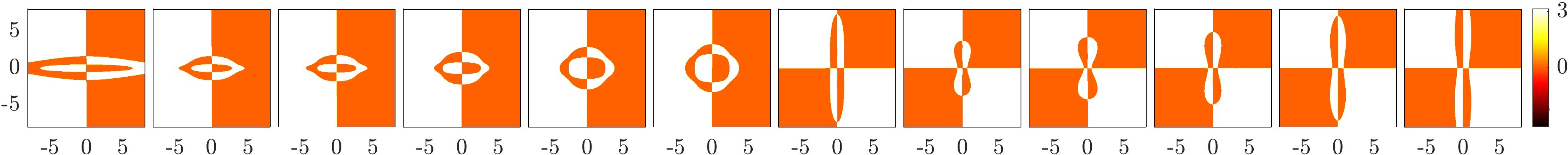}
\includegraphics[width=\textwidth]{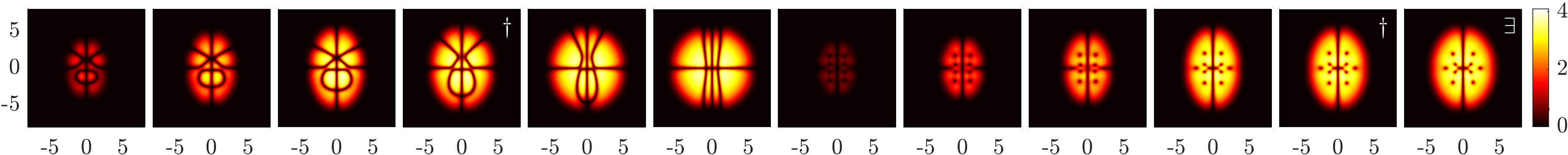}
\includegraphics[width=\textwidth]{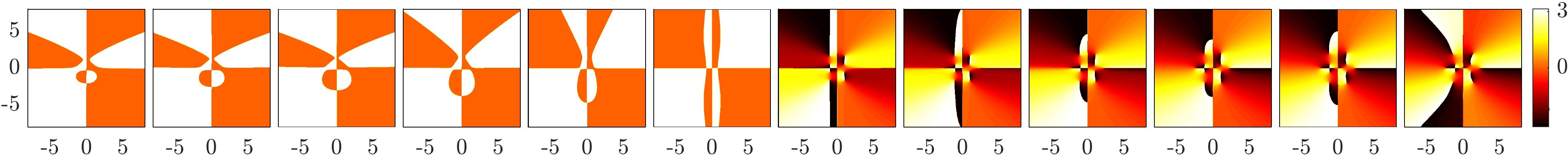}
\includegraphics[width=\textwidth]{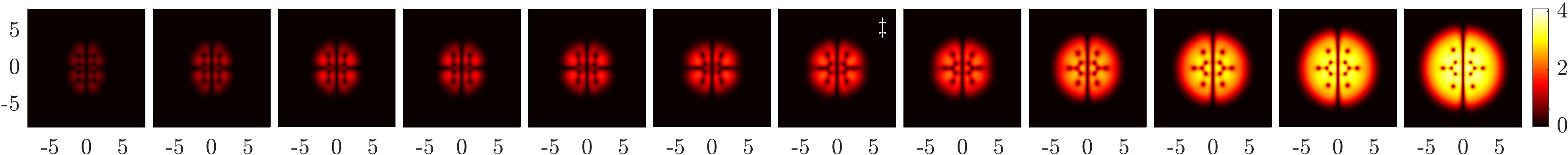}
\includegraphics[width=\textwidth]{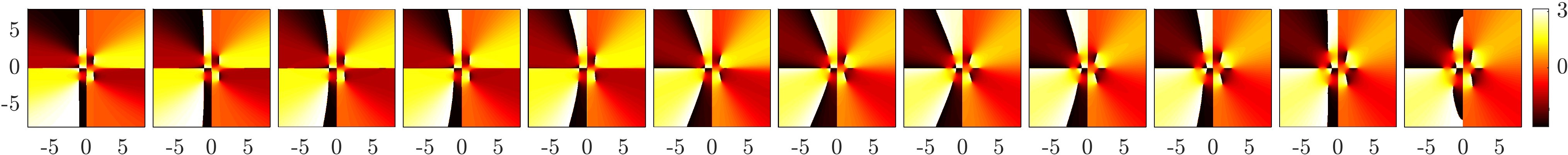}
\caption{
Solitary waves continued from the $\mu_0=6.25, 6.75$ linear degenerate sets of the $\kappa=2/3$ trap. The first set shows the DS05 and DS22 states at $\mu=7, 11, 16$ and $\omega_x=1.3, 1.1, 1$. The second set depicts the U3O state at $\mu=7, 11, 16$ and $\omega_x=1.3, 1.2, 1$, and the VX10 state at $\mu=7, 11, 14, 16$ and $\omega_x=1.3, 1$. The third set shows the DS14 and DS31 states at $\mu=8, 12, 16$ and $\omega_x=1.3, 1.1, 1$. The fourth set depicts the DSU2O soliton at $\mu=8, 12, 16$ and $\omega_x=1.3, 1.1, 1$, and the DSVX8 state at $\mu=7, 9, 12, 16$ and $\omega_x=1.4, 1.275$. Here, the DSVX8 state reaches an existence boundary, however, this state can be continued into the isotropic trap in the near-linear regime, as shown in the final set at $\mu=7$. The states of $\omega_x=1.5, 1.4, 1.225, 1.224, 1.1, 1$ are illustrated. Finally, the state in the isotropic trap is continued further in the chemical potential, and the states of $\mu=7.5, 8, 10, 12, 14, 16$ are sketched.
}
\label{DS05Aa}
\end{figure*}

The tenth linear degenerate set has two basis states \statem{05} and \statem{22} with eigenenergy $\mu_0=6.25$. We have identified a total of four states, and their respective ULSs are:
\begin{align}
    \varphi_{\mathrm{DS05}}^0 &= \state{05}, \\
    \varphi_{\mathrm{DS22}}^0 &= \state{22}, \\
    \varphi_{\mathrm{U3O}}^0 &= 0.6021\state{05} -0.7984\state{22}, \\
    \varphi_{\mathrm{VX10}}^0 &= 0.6667\state{05}+ 0.7453i\state{22}.
\end{align}
These coefficients are estimated using the numerically exact states at $\mu=6.29$. The continuation of these states is illustrated in Fig.~\ref{DS05Aa}.

The continuation of the DS05 state is qualitatively similar to that of the DS04 state above in Fig.~\ref{DS04Aa}, except that there is an additional horizontal dark soliton stripe. However, they also differ in details, note that the inner RDS is quite elongated along the $y$ axis in the isotropic trap, i.e., the rotational symmetry of the RDS2 by $\pi/2$ therein is broken due to the presence of the dark soliton stripe. Interestingly, this state also differs from its counterpart of RDS2XDS in \cite{Wang:LLC}. The dark soliton stripe passes through the rounded corner of the outer and inner deformed ring dark solitons, respectively.

The DS22 state quickly forms four lobes of droplets along the $x$ and $y$ axes, while the rest of the droplets connect together. This feature becomes clear in the TF regime. As it is continued into the isotropic trap, it gradually morphs into the double $\phi$ soliton instead of the DS22 state \cite{Wang:LLC}. Note that dark soliton reconnections happen at $\omega_x \approx 1.11$ such that the central droplet becomes an isolated one in the isotropic trap.

The U3O soliton has three U solitons and a small RDS in the small-density regime. As the chemical potential increases, the three U solitons connect together at $\mu \approx 7.5$ in our spatial horizon, forming a single complicated dark soliton filament. As $\omega_x$ decreases, two pairs of dark soliton reconnections occur at $\omega_x\approx1.25$ and $1.18$, respectively, leading to the waveform in the isotropic trap. This final state is like the final UO state of Fig.~\ref{DS03Aa} in coexistence with a RDS, and it is also like an isomer of the double $\phi$ soliton.

The VX10 state has an array of $5\times2$ vortices in the weakly-interacting regime, and the vortex charges are $1, -1, -1, -1, 1$ on the left side and the right ones have the opposite charges. As the chemical potential increases, four vortices of opposite charges are quickly induced at $\mu \approx 6.32$ along the four density depletion regions in our spatial horizon. Similarly, two additional vortices of alternating charge are induced at $\mu\approx7.9$ along the $x$ axis, yielding a VX16 state. 
%and these induced vortices have alternating charges. 
The former four vortices are induced into the condensate at $\mu \approx 12.5$. Next, two pair creations occur at $\mu\approx14.4$ in the central region such that the central vortices have alternating charges, leading to a VX20 state. Finally, the two vortices along the $x$ axis are also induced into the condensate at $\mu \approx 15.0$. The trap continuation is pretty robust, despite that the wave structure is quite complicated.

The eleventh linear degenerate set has two basis states \statem{14} and \statem{31} with eigenenergy $\mu_0=6.75$. We have identified a total of four states, and their respective ULSs are:
\begin{align}
    \varphi_{\mathrm{DS14}}^0 &= \state{14}, \\
    \varphi_{\mathrm{DS31}}^0 &= \state{31}, \\
    \varphi_{\mathrm{DSU2O}}^0 &= 0.7573\state{14} -0.6531\state{31}, \\
    \varphi_{\mathrm{DSVX8}}^0 &= 0.7254\state{14} +0.6884i\state{31}.
\end{align}
These coefficients are estimated using the numerically exact states at $\mu=6.79$. The continuation of these states is depicted in Fig.~\ref{DS05Aa}. It is interesting to compare with the states from the eighth linear degenerate set as sketched in Fig.~\ref{DS04Aa}.
    
The continuation of the DS14 state is very similar to that of the DS04 state in Fig.~\ref{DS04Aa} except that there is an additional vertical dark soliton stripe, compare also with the continuation of the DS05 state above. 
%As such, the two RDSs are less symmetric. 
Interestingly, the DS14 and DS05 states become identical in the isotropic trap despite that they have different orientations, to our knowledge. 
The DS31 state is like a composition of the DS8 and the XDS2 states in the TF regime. As $\omega_x$ decreases, it gradually deforms into the regular DS31 state of the isotropic trap \cite{Wang:LLC}.

The DSU2O state has a vertical dark soliton stripe, two U solitons, and a RDS. The chemical potential continuation is pretty robust. As $\omega_x$ decreases, the state gradually morphs into the regular DS31 state in the isotropic trap \cite{Wang:LLC}. Here, the dark soliton filaments deform gradually, and the dark soliton connections only happen in the vicinity of $\omega_x=1$.

The DSVX8 state has a vertical dark soliton stripe and an array of $4 \times 2$ vortices from a complex mixing of the \statem{14} and \statem{31} states in the near-linear regime. In this way, the vortices are of alternating charge except the two central vertical pairs. Therefore, two edge vortices of alternating charge are quickly induced along $y=0$ and they appear at $\mu \approx 6.9$ in our spatial horizon. %and each has the opposite charge to their two neighbouring central vortices. 
These two vortices are induced into the condensate around $\mu=12$. In addition, two vortices bifurcate in the center of the condensate, such that the neighbouring vortices are of alternating charge. As it is continued into the isotropic trap, the two vortical stripes bend outwards slightly until reaching an existence boundary at $\omega_x \approx 1.275$. This critical value diminishes only slightly as $\mu$ decreases. However, a successful continuation is found at $\mu=7$ in the near-linear regime. This evolved state features two edge vortices, as discussed above. As $\omega_x$ decreases, two additional edge vortices are induced and two pair annihilations happen at $\omega_x \approx 1.224$, yielding again a DSVX8 state with two strong density depletion regions. Next, two pair creations happen therein at $\omega_x \approx 1.006$, creating a DSVX12 state. Finally, the chemical potential continuation of this state in the isotropic trap is pretty robust. 

\begin{figure*}[t]
\includegraphics[width=\textwidth]{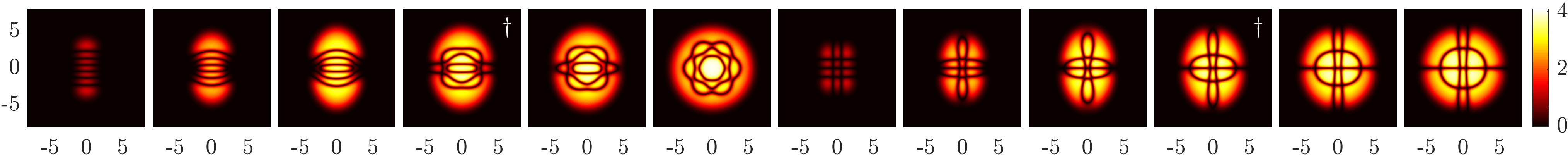}
\includegraphics[width=\textwidth]{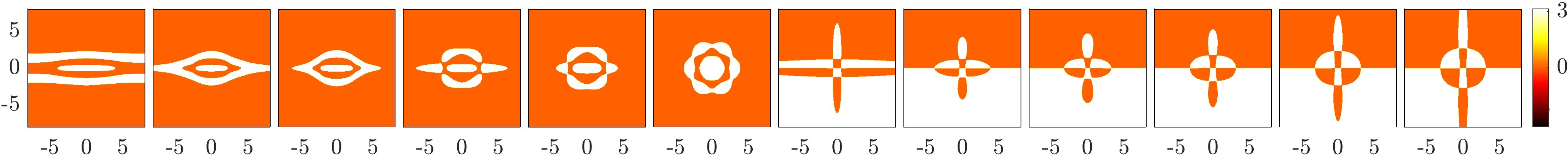}
\includegraphics[width=\textwidth]{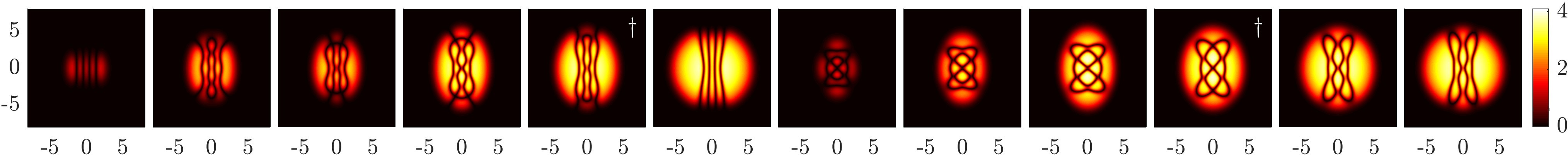}
\includegraphics[width=\textwidth]{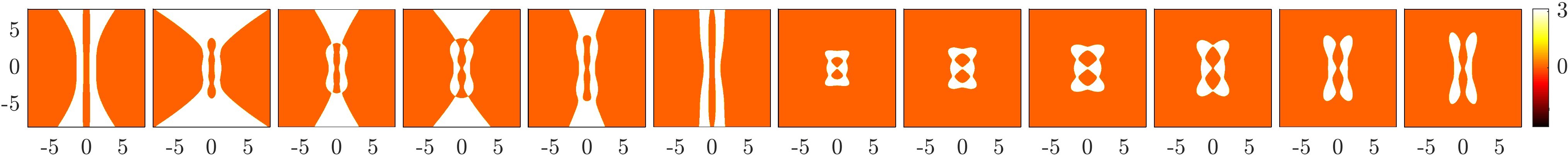}
\includegraphics[width=\textwidth]{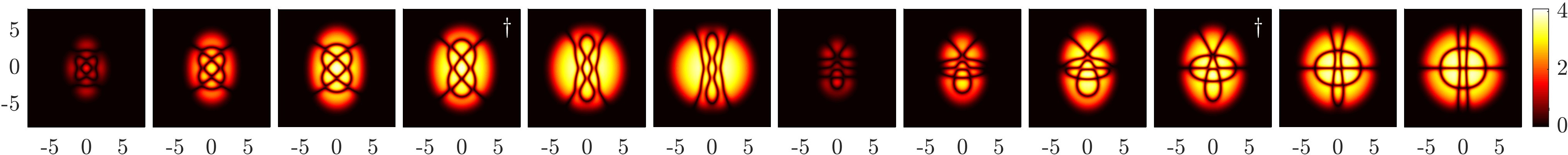}
\includegraphics[width=\textwidth]{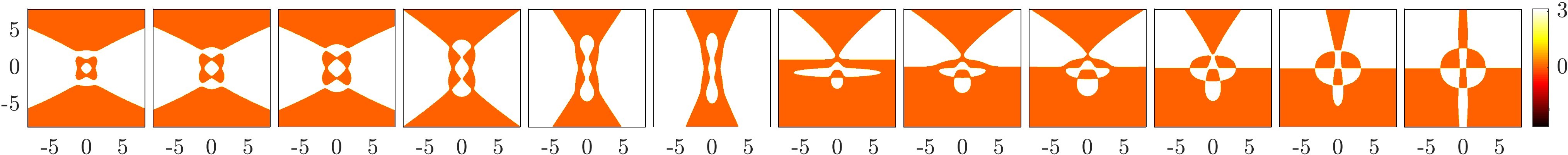}
\includegraphics[width=\textwidth]{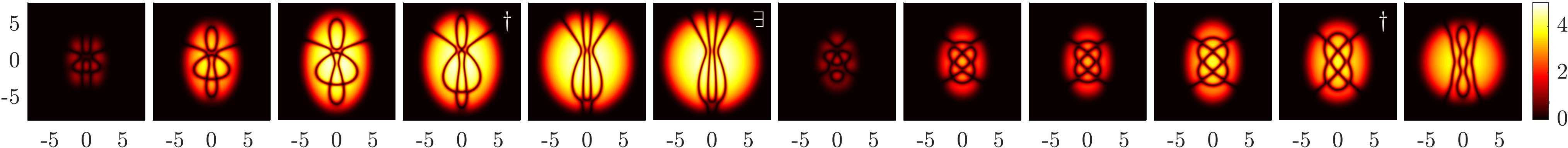}
\includegraphics[width=\textwidth]{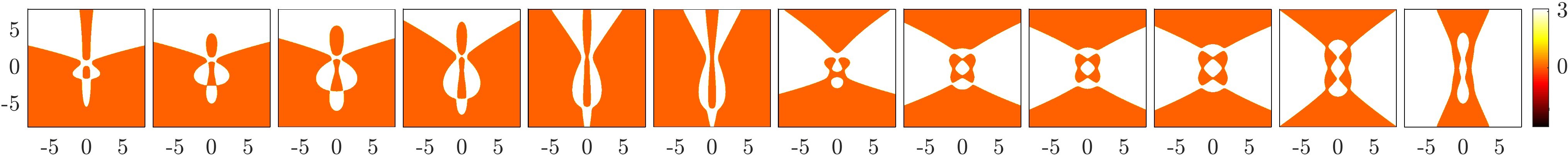}
\caption{
Solitary waves continued from the $\mu_0=7.25$ linear degenerate set of the $\kappa=2/3$ trap, part one. The first set shows the DS06 and DS23 states at $\mu=8, 12, 16$ and $\omega_x=1.3, 1.2, 1$, and $\omega_x=1.3, 1.1, 1$, respectively. The second set depicts the DS40 state at $\mu=8, 12.088, 12.09, 16$ and $\omega_x=1.3, 1$, and the O3 state at $\mu=8, 12, 16$ and $\omega_x=1.3, 1.1, 1$. The third set illustrates the U2O2 and U4O states at $\mu=8, 12, 16$ and $\omega_x=1.3, 1.1, 1$. The fourth set sketches the $\Psi$2 state at $\mu=8, 16, 24$ and $\omega_x=1.3, 1.1, 1.043$, and the VWO state at $\mu=8, 11.594, 11.596, 16$ and $\omega_x=1.3, 1$. The $\Psi$2 state can be continued into the isotropic trap at $\mu=33$, see the text for details.
}
\label{DS06Aa}
\end{figure*}

\begin{figure*}[t]
\includegraphics[width=\textwidth]{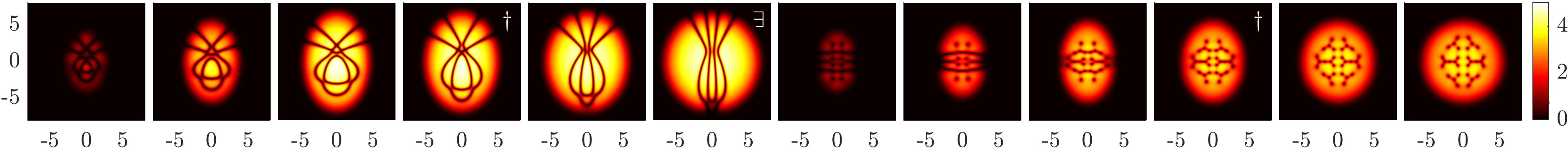}
\includegraphics[width=\textwidth]{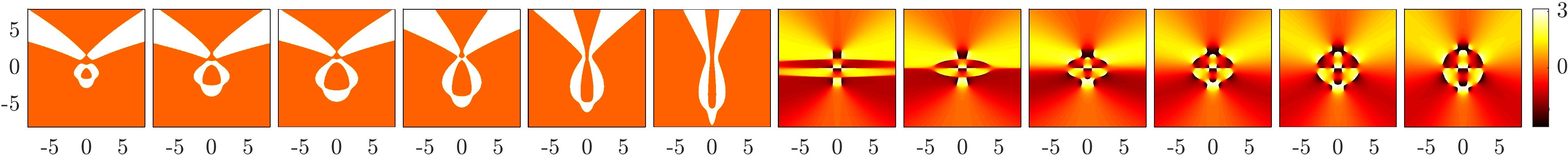}
\includegraphics[width=\textwidth]{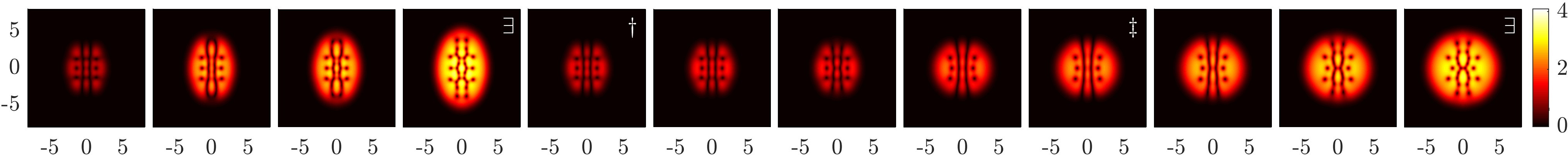}
\includegraphics[width=\textwidth]{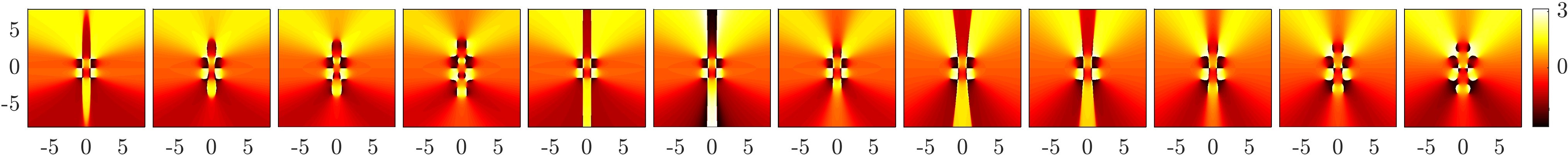}
\includegraphics[width=\textwidth]{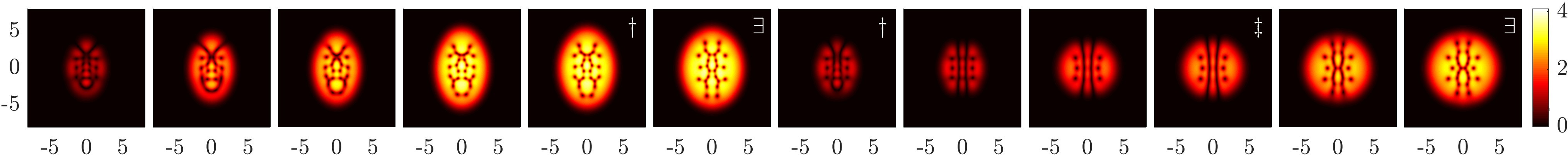}
\includegraphics[width=\textwidth]{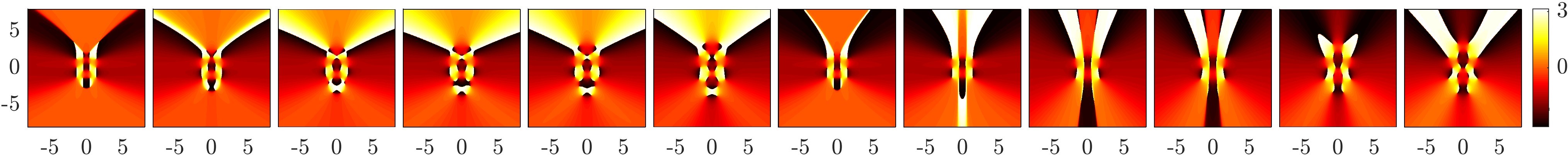}
\includegraphics[width=\textwidth]{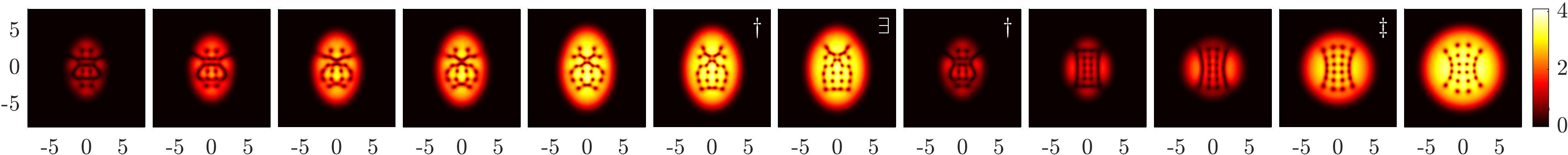}
\includegraphics[width=\textwidth]{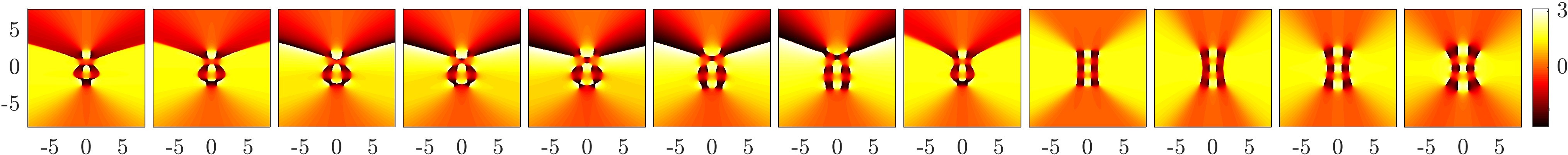}
\caption{
Solitary waves continued from the $\mu_0=7.25$ linear degenerate set of the $\kappa=2/3$ trap, part two. The first set shows the VWO2 state at $\mu=8, 16, 24$ and $\omega_x=1.35, 1.2, 1.043$, and the VX14 state at $\mu=8, 12, 16$ and $\omega_x=1.3, 1.1, 1$. The VWO2 state can be continued into the isotropic trap at $\mu=33$, see the text for details. The second set depicts the VX16 state at $\mu=8, 11.5, 12.5, 15.228$, reaching an existence boundary which also affects the trap continuation. However, the state can be continued into the isotropic trap at $\mu=8$ and the states at $\omega_x=1.4, 1.3225, 1.3220, 1$ are shown. Finally, the state is further continued in the chemical potential, and the states at $\mu=9, 10, 11, 12.648$ are depicted, reaching again an existence boundary.  
The third set illustrates the VX20 state at $\mu=8, 11, 13, 16$ and $\omega_x=1.4, 1.2595$, reaching an existence boundary. However, the state can be continued into the isotropic trap at $\mu=8$ and the states at $\omega_x=1.43, 1.3815, 1$ are shown. Finally, the state is further continued in the chemical potential, and the states at $\mu=9, 11, 12.64$ are depicted, reaching again an existence boundary. 
The fourth set sketches the VX20b state at $\mu=8, 10, 12.5, 13.5, 16$ and $\omega_x=1.4, 1.339$, reaching an existence boundary. However, the state can be continued into the isotropic trap at $\mu=8$ and the states at $\omega_x=1.4, 1.3, 1$ are shown. Finally, the state is further continued in the chemical potential, and the states at $\mu=12, 16$ are depicted.
}
\label{DSVWO2Aa}
\end{figure*}

\begin{figure*}[t]
\includegraphics[width=\textwidth]{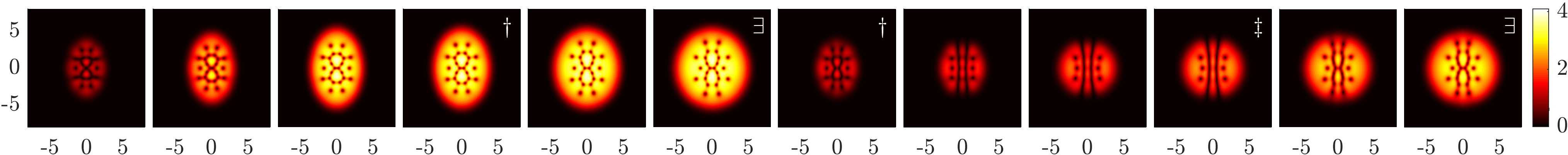}
\includegraphics[width=\textwidth]{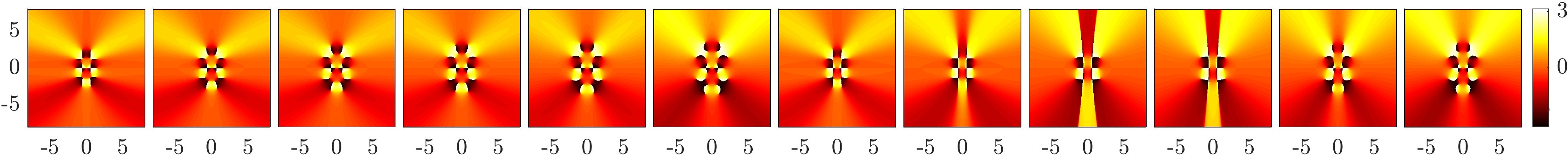}
\includegraphics[width=\textwidth]{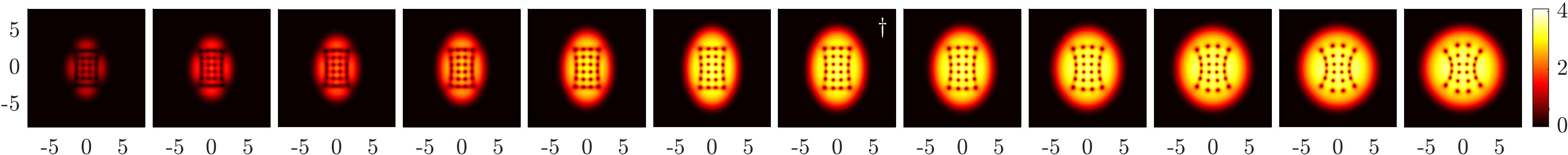}
\includegraphics[width=\textwidth]{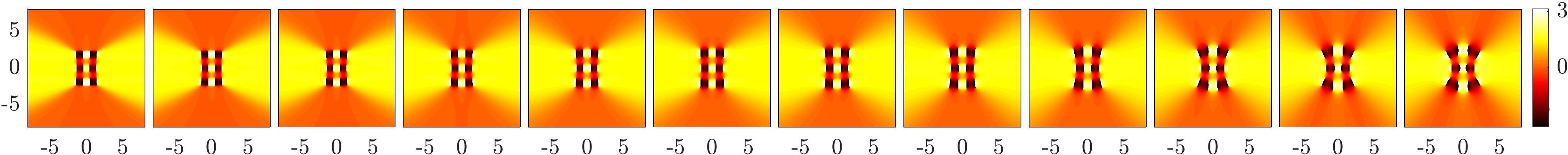}
\includegraphics[width=\textwidth]{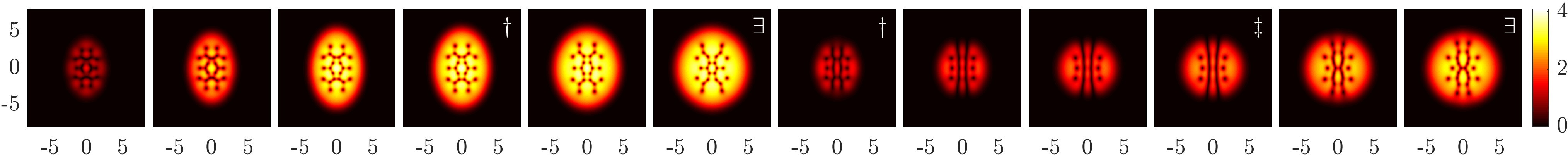}
\includegraphics[width=\textwidth]{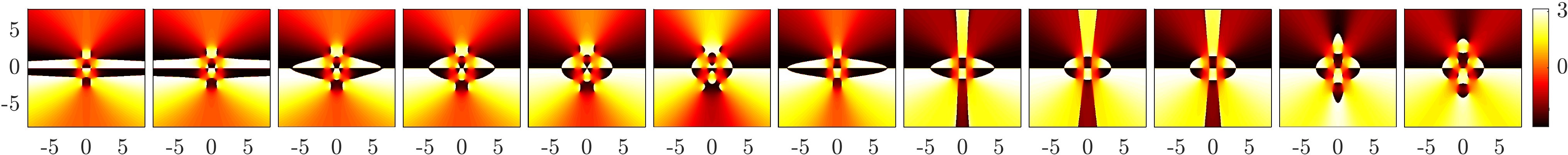}
\caption{
Solitary waves continued from the $\mu_0=7.25$ linear degenerate set of the $\kappa=2/3$ trap, part three.
The first set shows the VX22 state at $\mu=8, 12, 16$ and $\omega_x=1.4, 1.2, 1.0755$, reaching an existence boundary. However, the state can be continued into the isotropic trap at $\mu=8$ and the states at $\omega_x=1.4, 1.3, 1$ are shown. Finally, the state is further continued in the chemical potential, and the states at $\mu=9, 11, 12.64$ are depicted, reaching again an existence boundary. The second set illustrates the VX24 state at $\mu=8, 9, 10, 12, 14, 16$ and $\omega_x=1.4, 1.3, 1.2, 1.1, 1.05, 1$.
The third set sketches the VX24b state at $\mu=8, 12, 16$ and $\omega_x=1.4, 1.2, 1.089$, reaching an existence boundary. However, the state can be continued into the isotropic trap at $\mu=8$ and the states at $\omega_x=1.35, 1.2, 1$ are shown. Finally, the state is further continued in the chemical potential, and the states at $\mu=9, 11, 12.64$ are depicted, reaching again an existence boundary.
}
\label{VX22Aa}
\end{figure*}

The twelfth linear degenerate set has three basis states \statem{06}, \statem{23} and \statem{40} with eigenenergy $\mu_0=7.25$. Again, we reach the degeneracy of $3$ at the level of quantum number $6$, and the pattern formation is substantially richer. We have identified the following solitary waves:
\begin{align}
    \varphi_{\mathrm{DS06}}^0 &= 0.99997\state{06} +0.00018406\state{40}, \\
    \varphi_{\mathrm{DS23}}^0 &= \state{23}, \\
    \varphi_{\mathrm{DS40}}^0 &= -0.00071352\state{06} +0.99996\state{40}, \\
    \varphi_{\mathrm{O3}}^0 &= 0.7408\state{06} +0.6717\state{40}, \\
    \varphi_{\mathrm{U2O2}}^0 &= 0.7415\state{06} -0.6709\state{40}, \\
    \varphi_{\mathrm{U4O}}^0 &= 0.5512\state{06} -0.8342\state{23} +0.0170\state{40}, \\
    \varphi_{\Psi\mathrm{2}}^0 &= 0.0164\state{06} -0.8728\state{23} +0.4878\state{40}, \\
    \varphi_{\mathrm{VWO}}^0 &= 0.5967\state{06} -0.5945\state{23} -0.5389\state{40}, \\
    \varphi_{\mathrm{VWO2}}^0 &= 0.6077\state{06} -0.5742\state{23} +0.5485\state{40}, \\
    \varphi_{\mathrm{VX14}}^0 &= 0.6533\state{06} +0.7571i\state{23} +0.0040\state{40}, \\
    \varphi_{\mathrm{VX16}}^0 &= 0.0030\state{06} +0.7815i\state{23} +0.6238\state{40}, \\
    \varphi_{\mathrm{VX20}}^0 &= 0.5026\state{06} -(0.6077-0.0003i)\state{23} \nonumber \\ 
    &-(0.0065+0.6148i)\state{40}, \\
    \varphi_{\mathrm{VX20b}}^0 &= 0.6464\state{06} +(0.0067-0.6232i)\state{23} \nonumber \\ 
    &-(0.0027-0.4401i)\state{40}, \\
    \varphi_{\mathrm{VX22}}^0 &= 0.5411\state{06} +0.6839i\state{23} +0.4893\state{40}, \\
    \varphi_{\mathrm{VX24}}^0 &= 0.7315\state{06} +(0.0009+0.6818i)\state{40}, \\
    \varphi_{\mathrm{VX24b}}^0 &= 0.5410\state{06} -0.6844i\state{23} -0.4887\state{40}.
\end{align}
These coefficients are estimated using the numerically exact states at $\mu=7.29$. The continuation of these states is illustrated in Figs.~\ref{DS06Aa}-\ref{VX22Aa}.

The DS06 state is largely from the \statem{06} state with a small mixing from the \statem{40} state in the near-linear regime. Therefore, the dark solitons slightly deform, e.g., the central two dark solitons form a closed loop. As the chemical potential increases, they form three deformed concentric loops in the TF regime. As it is continued into the isotropic trap, the two outer rings undergo dark soliton reconnections at $\omega_x \approx 1.39$, yielding a fragmented structure. However, these dark solitons then reconnect back into ring structures at $\omega_x \approx 1.23$. Interestingly, the final state features a six-fold symmetry, and it is a solitary wave isomer of the polar RDS3 state \cite{Wang:LLC}. 

The continuation of the DS23 state is pretty robust. It approximately forms two loops with a horizontal dark soliton stripe in the TF regime. As $\omega_x$ decreases, it gradually morphs into a double $\phi$ solition in coexistence with the dark soliton stripe, perhaps we can call it the $\phi_{xyy}$ state. This final state has a linear limit in the isotropic trap \cite{Panos:DC1}. There appears to be a rich family of such $\phi$-related wave patterns, e.g., the triple $\phi$ state also exists \cite{Panos:DC1}.

The inner two dark soliton stripes of the DS40 state form a closed loop in the near-liner regime, however, it quickly opens around $\mu=8$. Next, the dark soliton filaments start to deform above $\mu \approx 11$, and the inner two dark soliton stripes form a closed loop again at $\mu \approx 11.7$. Then, the DS40 makes a sudden transition between $\mu=12.088$ and $\mu=12.09$. In this process, the outer dark soliton stripes approximately form a peanut shaped dark soliton loop, which is very similar to the DS20 to DS8 transition shown in Fig.~\ref{DS06a}. As it is continued into the isotropic trap, the filaments become smoother and the inner loop reopens around $\omega_x=1.1$, forming a regular DS40 state. Interestingly, this regular DS40 state was not obtained in the isotropic trap \cite{Wang:LLC}, but it is indirectly found here.

The O3 soliton has two small RDSs and a large deformed RDS around them in the low-density regime, and the chemical potential continuation of this state is pretty robust. As it is continued into the isotropic trap, two dark soliton reconnections occur. First, the two inner RDSs form a single loop at $\omega_x \approx 1.37$, and subsequently two peanut structures form at $\omega_x \approx 1.27$. This final state is identical to the DS40 state of the isotropic trap \cite{Wang:LLC}, to our knowledge.

The U2O2 soliton has two deformed RDSs in the middle with two U solitons on the elongated sides. The chemical potential continuation is quite robust. As it is continued into the isotropic trap, two dark soliton reconnection processes occur with four outer ones and two inner ones at $\omega_x\approx1.40, 1.38$, respectively. The final configuration is like a DS20 state with a central peanut loop in the isotropic trap.

The U4O state has a deformed RDS, two deformed U solitons forming a loop structure, and two U solitons on the sides, and the chemical potential continuation is quite robust. However, the trap continuation is somewhat complicated, as it has a series of dark soliton deformations and reconnections. We should not discuss the details here, as they are not fully essential to characterize the main features. Roughly speaking, the two U solitons on the top gradually open, and all the dark soliton filaments form an intermediate state, which approximately has an asymmetric U state, a RDS and a horizontal dark soliton stripe, at $\omega_x \approx 1.3$. Finally, the asymmetric U state gradually deforms into a DS20 state approximately, yielding the $\phi_{xyy}$ waveform in the isotropic trap. This final state is identical to the above DS23 counterpart, to our knowledge.

The chemical potential continuation of the $\Psi$2 state is quite robust, except that there are two minor dark soliton reconnection processes at $\mu\approx12.5$ and $19.0$, respectively, see the phase profiles for details. As it is continued into the isotropic trap, the outer filament becomes less curved and its upper part moves closer to the $y$ axis. In addition, the inner two loops reconnect into a single loop at $\omega_x \approx 1.27$. The final configuration is like a parity symmetry-breaking U2 state, and there is an existence boundary at $\omega_x \approx 1.043$. This critical frequency decreases slowly with increasing chemical potential, and this state is successfully continued into the isotropic trap at $\mu=33$. As the pertinent states are very similar in structure, we should not present the process at this quite large chemical potential in detail for simplicity.

The VWO state has a quite complicated structure in the near-linear regime. Two dark soliton reconnection processes occur at $\mu \approx 11.5$, yielding a waveform resembles a less symmetric U2O2 state. Then the state makes a transition between $\mu=11.594$ and $11.596$, and it becomes identical to that of the U2O2 wave pattern, to our knowledge. Naturally, the subsequent evolution in the chemical potential and the trap is also the same, and we should not discuss it again here. 

The VWO2 state contains two U solitons very close to each other and also two highly deformed RDSs. The chemical potential continuation is very robust. As $\omega_x$ decreases, two dark soliton reconnection processes take place at $\omega_x\approx1.38$ and $1.31$, respectively. The final state is very similar to the above $\Psi$2 counterpart. Interestingly, they encounter the same existence boundary at $\omega_x\approx1.043$ despite that they remain slightly different before they cease to exist, to our knowledge. This feature persists for a few other chemical potentials that we investigated. Similarly, the state is successfully continued into the isotropic trap at $\mu=33$, and again it is very similar to but slightly different from the $\Psi$2 state. Because the relevant waveforms are qualitatively similar, we should not present them for simplicity. 
%We changed the domain box parameter from $8$ to $10$ for both states.

The VX14 state is largely from a complex mixing of the \statem{06} and \statem{23} states, yielding an array of $6\times2$ vortices, however, there are also two edge vortices along $y=0$ in the near-linear regime. As the chemical potential increases, four pair creations occur on the sides of the two major density depletion regions at $\mu \approx 8.4$, leading to a VX22 state. Next, two pair creations take place in the middle of the two density depletion regions at $\mu \approx 13.2$, yielding a VX26 state in the TF regime. As it is continued into the isotropic trap, two pair creations occur along $y=0$ at $\omega_x \approx 1.41$, forming a VX30 state with four Y clusters and two dog bone structures.

The VX16 state has two density depletion regions and each contains $5$ vortices in the near-linear regime. The neighbouring vortices have alternating charge except the three central pairs on each side. As the chemical potential increases, the two central vortices undergo elongation and pair creation at $\mu \approx 11.1$, yielding a VX20 state. Next, four pair creations occur next to the four edge vortices at $\mu \approx 12.0$, leading to a VX28 state. The neighbouring vortices alternate in charge in this final state. Interestingly, we find an upper critical chemical potential at $\mu_c \approx 15.228$, despite that the state has a linear limit in this potential trap. This existence boundary also affects the trap continuation, e.g., the critical frequency is only $\omega_x \approx 1.499$ at $\mu=15.2$ and it declines very slowly with decreasing chemical potential. However, the VX16 state is successfully continued into the isotropic trap at $\mu=8$ in the low-density regime. Here, the four edge vortices are gradually ejected out of the condensate at $\omega_x \approx 1.41$, but then they are suddenly induced into the condensate again between $\omega_x \approx 1.3225$ and $1.3220$. As such, the final state in the isotropic trap remains very similar in structure. We subsequently continue this state further in chemical potential in the isotropic trap. As the chemical potential increases, four pair creations occur at $\mu \approx 10.3$ next to the central vortices, forming a VX24 state. Finally, this state again reaches an existence boundary at $\mu_c \approx 12.648$.

The VX20 state is largely from a complex mixing of the U4O and DS40 states in the near-linear regime. As the chemical potential increases, the two edge vortices are gradually ejected out of the condensate and they disappear in our spatial horizon at $\mu \approx 10.6$, forming a VX18 state. Next, two pair creations occur in the lower part of the condensate at $\mu \approx 12.5$, yielding two Y clusters therein and consequently a VX22 state. Similarly, two pair creations also occur in the top part of the condensate at $\mu \approx 13.1$, leading to two Y clusters therein and consequently a VX26 state. As it is continued into the isotropic trap, the vortices adjust themselves until reaching an existence boundary at $\omega_x \approx 1.2595$. The final state is very similar to the above VX28 state, which in turn is from the VX16 state, except that the vortex dipole at the top is missing. The critical frequency only declines slowly with decreasing chemical potential, however, a successful continuation is found at $\mu=8$. As $\omega_x$ decreases, the two edge vortices at the top are gradually ejected out of the condensate and they disappear in our spatial horizon at $\omega_x \approx 1.44$, generating a VX18 state. Soon, two pair annihilations occur in the top density depletion region at $\omega_x \approx 1.41$, leading to a VX14 state. Meanwhile, the density depletion region at the bottom gradually extends outwards and consequently another two vortices are ejected out of the condensate at $\omega_x \approx 1.383$, forming a VX12 structure. Next, the state becomes more symmetric and turns into its counterpart of the evolved VX16 state (see above) at $\omega_x \lesssim 1.3815$, to our knowledge. Therefore, the subsequent evolution and also the further continuation in the chemical potential in the isotropic trap are identical, and we should not discuss them here for simplicity.

The VX20b state is largely from a complex mixing of the $\Psi$2 and DS06 states in the near-linear regime. As the chemical potential increases, a pair creation occurs slightly above the central region at $\mu \approx 8.5$, leading to a VX22 state. Next, the edge VX2 are ejected out and they disappear in our spatial horizon at $\mu \approx 11.2$ and a pair annihilation occurs below the top vortex dipole at $\mu \approx 12.0$, yielding a VX18 state. Two pair creations follow at $\mu \approx 12.9$ in the density depletion region, forming another VX22 state. Interestingly, another pair creation occurs slightly above the central region at $\mu \approx 13.6$, forming four Y clusters of alternating charge. Finally, a pair creation happens in the bottom of the condensate, leading to a VX26 state. As $\omega_x$ decreases, the vortices adjust their positions before reaching an existence boundary at $\omega_x \approx 1.339$. The final configuration is like a $4 \times 4$ vortex lattice with a vortex dipole and two Y structures. Again, the critical frequency only declines slowly with decreasing chemical potential, but a successful continuation is found at $\mu=8$. First, a pair creation quickly happens slightly above the central region at $\omega_x \approx 1.494$. Similarly, a pair creation occurs at the bottom of the condensate at $\omega_x \approx 1.32$, yielding a $6\times4$ miniature lattice. The vortices slightly adjust themselves and then the waveform becomes quite symmetric at $\omega_x \approx 1.306$. Finally, we continue this lattice further in the chemical potential. Two pair annihilations occur in the central outer stripes at $\mu \approx 8.1$, but they are quickly followed by two pair creations therein at $\mu \approx 8.8$ restoring the lattice structure. The subsequent continuation is robust. Note that the ``lattice planes'' are quite bent in the isotropic trap.

The VX22 state is largely from a complex mixing of the O3 and DS23 states in the near-linear regime. As the chemical potential increases, a pair creation occurs in the central region at $\mu \approx 9.9$, yielding a VX24 state. As it is continued into the isotropic trap, the vortices adjust their positions before reaching an existence boundary at $\omega_x \approx 1.0755$. The critical frequency only diminishes slowly with decreasing chemical potential, but a successful continuation is found at $\mu=8$. As $\omega_x$ decreases, a pair creation quickly occurs in the central region at $\omega_x \approx 1.498$, yielding a VX24 state. Next, four pair annihilations happen surrounding this vortex dipole at $\omega_x \approx 1.39$, leading to a VX16 state. Interestingly, the counterparts of the evolved VX16 and VX20 states in Fig.~\ref{DSVWO2Aa} and the state here become identical from $\omega_x\approx1.322$, including the subsequent evolutions, to our knowledge. Therefore, we should not discuss them again for simplicity.

The VX24 state is largely from a complex mixing of the \statem{06} and \statem{40} states in the near-linear regime. It forms a miniature $6 \times 4$ vortex lattice of alternating charge, and its continuation is quite robust. The vortex lattice somewhat bends as it is continued into the isotropic trap. Note that we also obtained this wave pattern in the study of the VX20b state, and the final states are identical, to our knowledge.

The VX24b state is largely from a complex mixing of the U2O2 and \statem{23} states in the near-linear regime. It has $6$ Y clusters of alternating charge, and its chemical potential continuation is quite robust. As it is continued into the isotropic trap, the vortices adjust their positions until reaching an existence boundary at $\omega_x\approx1.0890$. Interestingly, we have obtained two similar states before, one has an additional vortex dipole at the bottom and the other has two addition vortex dipoles at the top and the bottom; see the evolution of the above VX20 and VX16 states, respectively. The critical frequency only diminishes very slowly as the chemical potential decreases, however, a successful continuation is obtained at $\mu=8$. As $\omega_x$ decreases, two vortex merging processes occur in the center at $\omega_x \approx 1.38$, leading to a VX20 state. Next, four pair annihilations take place at the edge of the density depletion region. However, four vortices are quickly induced therein at the transition between $\omega_x=1.3225$ and $1.3220$. We have seen this transition several times. In summary, this state and its counterparts of the evolved VX16, VX20, and VX22 states above all become identical from $\omega_x\approx1.322$, to our knowledge. In addition, the pertinent evolution of the VX22 state is smooth, while the other three states exhibit transitions. In this way, the subsequent evolutions are already known, and we should not discuss them again for simplicity.

\begin{table}[t]
\caption{
Number of distinct solitary waves $\Omega$ identified in the near-linear regime by the linear limit continuation method. Here, $g$ is the degeneracy of the linear degenerate set of $\kappa$ and $\mu_0$, $\Omega_r$ and $\Omega_c$ are the numbers of real-valued dark soliton states and complex-valued states with vortices, respectively. The number of independent runs of the RRS and CRS is also summarized here for convenience. The linear degenerate sets of $\mu_0=1.25, 2.25, 2.75, 3.25, 3.75, 4.75$ of the $\kappa=2/3$ trap are nondegenerate. The data of $\kappa=1, 1/2$ are taken from \cite{Wang:LLC}.
\label{para}
}
\begin{tabular*}{\columnwidth}{@{\extracolsep{\fill}} l c c c c c c r}
\hline
\hline
$\kappa$ &$\mu_0$ &$g$ &$\Omega_r$ &$\Omega_c$ &$\Omega$ &RRS &CRS \\
\hline
$1$ &$1$ &$1$ &$1$ &$0$ &$1$ &- &- \\
\hline
$1$ &$2$ &$2$ &$1$ &$1$ &$2$ &$100$ &$100$ \\
\hline
$1$ &$3$ &$3$ &$3$ &$3$ &$6$ &$1000$ &$1000$ \\
\hline
$1$ &$4$ &$4$ &$4$ &$10$ &$14$ &$1000$ &$1000$ \\
\hline
$1$ &$5$ &$5$ &$9$ &$24$ &$33$ &$1000$ &$2000$ \\
\hline
$1/2$ &$1.5, 2.5$ &$1$ &$1$ &$0$ &$1$ &- &- \\
\hline
$1/2$ &$3.5$ &$2$ &$3$ &$1$ &$4$ &$1000$ &$1000$ \\
\hline
$1/2$ &$4.5$ &$2$ &$3$ &$1$ &$4$ &$1000$ &$1000$ \\
\hline
$1/2$ &$5.5$ &$3$ &$6$ &$6$ &$12$ &$1000$ &$1000$ \\
\hline
$1/3$ &$2, 3, 4$ &$1$ &$1$ &$0$ &$1$ &- &- \\
\hline
$1/3$ &$5$ &$2$ &$3$ &$1$ &$4$ &$1000$ &$1000$ \\
\hline
$1/3$ &$6$ &$2$ &$3$ &$1$ &$4$ &$1000$ &$1000$ \\
\hline
$1/3$ &$7$ &$2$ &$3$ &$1$ &$4$ &$1000$ &$1000$ \\
\hline
$1/3$ &$8$ &$3$ &$7$ &$7$ &$14$ &$1000$ &$1000$ \\
\hline
$2/3$ &$1.25, 2.25, ...$ &$1$ &$1$ &$0$ &$1$ &- &- \\
\hline
$2/3$ &$4.25$ &$2$ &$3$ &$1$ &$4$ &$1000$ &$1000$ \\
\hline
$2/3$ &$5.25$ &$2$ &$3$ &$1$ &$4$ &$1000$ &$1000$ \\
\hline
$2/3$ &$5.75$ &$2$ &$3$ &$1$ &$4$ &$1000$ &$1000$ \\
\hline
$2/3$ &$6.25$ &$2$ &$3$ &$1$ &$4$ &$1000$ &$1000$ \\
\hline
$2/3$ &$6.75$ &$2$ &$3$ &$1$ &$4$ &$1000$ &$1000$ \\
\hline
$2/3$ &$7.25$ &$3$ &$9$ &$7$ &$16$ &$1000$ &$1000$ \\
\hline
\hline
\end{tabular*}
\end{table}

Finally, we summarize the number of solitary waves bifurcated from the different linear degenerate sets that we have investigated so far in Table~\ref{para}. 
%The total number of solitary waves $\Omega$ grows very rapidly with increasing degeneracy $g$ for both potential traps. The exact scaling cannot be extracted accurately now, future work should study more excited states, but clearly it is strongly superlinear. 
The data of the $\kappa=1, 1/2$ traps are also presented here for completeness \cite{Wang:LLC}.
The basic trend that the number of real-valued waves, complex-valued waves, and the total number grow rapidly with increasing degeneracy $g$ holds for all the traps studied. More solitary waves are identified in the isotropic trap so far, however, it should be noted that relatively high degeneracy arises earlier in this trap. We expect the pattern formation of anisotropic traps is equally rich when $g$ is sufficiently large. This is promising as they do yield more wave patterns at the level of $g=2, 3$.
%The degeneracy grows slower for $\kappa=1/2$ because the neighbouring lattice points in a plane are further apart, however, more solitary waves appear to be available for the same degeneracy $g>1$. Furthermore, both the real-valued dark soliton states and the complex-valued states with vortices separately grow rapidly with increasing $g$. 
The number of independent runs of the real random searcher (RRS) and complex random searcher (CRS) is also summarized here for convenience, and see \cite{Wang:LLC} for details.

\section{Summary and outlook}
\label{co}

In this work, we applied the linear linear continuation method \cite{Wang:LLC} to two prototypical anisotropic traps of aspect ratio $\kappa=1/3$ and $2/3$ for systematically constructing solitary waves in two-dimensional Bose-Einstein condensates. 
Our efforts are largely successful. Many low-lying solitary waves (including parity symmetry-breaking ones) are identified in the near-linear regime and they are continued into the TF regime, and also into the isotropic trap when relevant. This is particularly true considering that we have ``only'' worked to the degeneracy $3$. It is evident that a good number of new wave patterns are found, such as the VX8a, VX12b, U3O, VX10, U2O2, VX14, and VX24 states to name a few, to our knowledge. In addition, the parametric connectivity of the low-lying states from $\kappa=1, 1/2, 1/3, 2/3$ is investigated when pertinent. This work further demonstrates that linear limit continuation is an effective method for systematically constructing numerical exactly solitary waves.

The linear limit continuation can be extended in several directions. First, it is natural to continue the efforts to study more harmonic traps, e.g., the low-lying aspect ratios of $\kappa=1/4, 3/4, 1/5, 2/5, 3/5, 4/5$ are interesting. Second, it is also relevant to study more excited linear degenerate sets. For example, preliminary simulations show that the $\mu_0=6$ linear degenerate set of the isotropic trap contains a large number of solitary waves, following the work of \cite{Wang:LLC}. Third, it is possible that different trap forms may encompass quite distinct wave patterns considering continuation chaos. In this vein, it is important to study a square or circular box potential \cite{BoxP,Angel:VX4}. It is also possible to explore quite generic potentials as the pertinent low-lying linear eigenvalues and eigenfunctions can be computed numerically.

It is particularly interesting to apply the method to the three-dimensional setting \cite{Ionut:VR,Wang:DSS,Panos:DC3,Brand_SV}, and also the two-component system \cite{Panos:DC2,Wang:VRB}. The theoretical framework in 3D remains essentially the same, despite that the computational work is considerably more challenging. Our preliminary result shows that more than $100$ solitary waves exist in the isotropic trap only upto the linear degenerate sets of quantum number $3$. The wave patterns of dark soliton surfaces and vortical filaments, as well as their chaotic evolution processes when relevant are very much worthy a systematic investigation. %It is highly interesting to extend the method to find vector solitary waves in multi-component BECs \cite{Panos:DC2,Wang:VRB}.
The framework of finding vector solitary waves in 1D has already been extensively explored \cite{Wang:DD,Wang:MDDD,Wang:DAD}. As a first step, it is interesting to combine the pertinent techniques of \cite{Wang:LLC,Wang:DD} to study the two-dimensional, two-component system. Recently, we find that the degeneracy further extends the 1D framework, it is also relevant to couple linear states from the same linear degenerate set, not just the distinct ones. For example, a vertical dark soliton stripe from \statem{10} can couple with a horizontal dark soliton stripe from \statem{01} in the isotropic trap, leading to a perpendicular dark-dark structure. Research efforts along some of these directions are currently in progress and will be reported in future publications.

%Finally, the detailed properties of the solitary waves are worth studying in their own rights theoretically, computationally, and experimentally. Here, one can investigate their equilibrium configurations at the reduced particle level \cite{Wang:AI,PK:DSVX}, their experimental implementations, stability properties, and dynamics. When the solitary waves are unstable, it is interesting to study their typical decay scenarios, and explore whether they can be fully stabilized by tuning the chemical potential and the trap or adding suitable external potential barriers.

\section*{Acknowledgments}
We gratefully acknowledge supports from the National Science Foundation of China under Grant No. 12004268, and the Fundamental Research Funds for the Central Universities, China. 
We thank the Emei cluster at Sichuan
University for providing HPC resources.

\bibliography{Refs}

\bibliographystyle{apsrevtitle}

\end{document}